\documentclass[journal]{IEEEtran}

\usepackage{amssymb}
\usepackage{amsmath}
 \usepackage{cite}
 \usepackage[font=footnotesize]{caption} 
\usepackage{bm}
\usepackage{float}
\usepackage{mathtools}
\usepackage{amsthm}
\usepackage{hyperref}       
\usepackage{url}            
\usepackage{xcolor}         
\usepackage{tikz}
\usepackage{pgfplots}
\usepackage{subcaption}
\usepackage[noend]{algpseudocode}

\pgfplotsset{compat=1.10}
\usepgfplotslibrary{fillbetween}

\pgfmathsetmacro\m{0.6} 
\pgfmathsetmacro\w{5} 

\newtheorem{theorem}{Theorem}

\newtheorem{definition}{Definition}
\newtheorem{proposition}{Proposition}
\newtheorem{remark}{Remark}

\newcommand{\norm}[1]{\left\lVert#1\right\rVert}
\makeatletter

\title{Offset-Symmetric Gaussians for Differential Privacy}

\author{Parastoo~Sadeghi,~\IEEEmembership{Senior Member,~IEEE,} and 
        Mehdi~Korki,~\IEEEmembership{Member,~IEEE}
\thanks{P. Sadeghi (contact author) is with the School of Engineering and Information Technology, the University of New South Wales, Canberra, Australia. Email: \texttt{p.sadeghi@unsw.edu.au}. M. Korki is with the Department of Telecommunications, Electrical, Robotics and Biomedical Engineering, Swinburne University of Technology, Melbourne, Australia. Email: \texttt{mkorki@swin.edu.au}.}}

\newcommand{\Lap}{\mathcal{L}}
\newcommand{\N}{\mathcal{N}}
\newcommand{\R}{\mathbb{R}}

\newcommand{\X}{\mathcal{X}}
\newcommand{\Y}{\mathcal{Y}}
\newcommand{\eps}{\varepsilon}
\newcommand{\del}{\delta}
\newcommand{\Prob}{\mathbb{P}}
\newcommand{\al}{\alpha}
\newcommand{\OSGT}{\mathcal{T}(0, V(m, \sigma^2))}
\newcommand{\OSGTK}{\mathcal{T}(0, V(m, \sigma^2)I_k)}

\begin{document}
\maketitle

\begin{abstract}
The Gaussian distribution is widely used in mechanism design for differential privacy (DP).~Thanks to its sub-Gaussian tail, it significantly reduces the chance of outliers when responding to queries.~However, it can only provide approximate $(\epsilon, \delta(\epsilon))$-DP. In practice, $\delta(\epsilon)$ must be much smaller than the size of the dataset, which may limit the use of the Gaussian mechanism for large datasets with strong privacy requirements.~In this paper, we introduce and analyze a new distribution for use in DP that is based on the Gaussian distribution, but has improved privacy performance. The so-called offset-symmetric Gaussian tail (OSGT) distribution is obtained through using the normalized tails of two symmetric Gaussians around zero. Consequently, it can still have sub-Gaussian tail and lend itself to analytical derivations. We analytically derive the variance of the OSGT random variable and the $\delta(\epsilon)$ of the OSGT mechanism. We then numerically show  that at the same variance, the OSGT mechanism can offer a lower $\delta(\epsilon)$ than the Gaussian mechanism. We extend the OSGT mechanism to $k$-dimensional queries and derive an easy-to-compute analytical upper bound for its zero-concentrated differential privacy (zCDP) performance. We analytically prove that at the same variance, the same global query sensitivity and for sufficiently large concentration orders $\alpha$, the OSGT mechanism performs better than the Gaussian mechanism in terms of zCDP.
\end{abstract}

\begin{IEEEkeywords}
Differential privacy, Gaussian mechanism, concentrated differential privacy, approximate differential privacy.
\end{IEEEkeywords}

\section{Introduction} 

Differential privacy (DP)~\cite{dworkoriginal2006, dwork2006calibrating, dwork2006our} is a widely-used statistical framework for protecting the privacy of individuals in datasets when they are to be shared with the public, used in training machine learning algorithms, or in responding to adaptive queries from adversarial data analysts.~Notable applications of DP include implementations by Google \cite{google}, Microsoft \cite{microsoft}, Apple \cite{apple}, and in deep learning~\cite{abadi2016deep}, \cite{GANobfuscator}, federated learning \cite{advances:FL}, \cite{Wei20} and the 2020 US Census \cite{uscensus}. In a typical implementation of DP, the privacy-preserving mechanism outputs a noisy version of the function of interest (i.e., the query) evaluated at the input dataset. Two most commonly used noise distributions are zero-mean Laplace and Gaussian \cite{dworkoriginal2006, dwork2006calibrating, dwork2006our}.

Until recently, the (discrete) Laplace mechanism was the primary contender for the upcoming public release of the 2020 US Census. However, on April 30, 2021, the US Census Bureau announced in \cite{uscensus:2021:discreteG} that the (discrete) Gaussian mechanism will be used in future releases. This  non-trivial change has been prompted by the lighter (sub-Gaussian) tail of the Gaussian distribution compared to the Laplace distribution. This property naturally carries over to their discrete counterparts and significantly reduces the probability of outliers, which is an important consideration as Census results can affect hundreds of millions of people. However, the price paid is that only approximate differential privacy is achievable, which is captured through an additive slack parameter $\delta$, in the so-called $(\epsilon, \delta)$-DP framework. For an acceptable performance on a dataset of size $n$, one must aim for $\delta \ll 1/n$. For the 2020 US Census release, $\delta < 10^{-10}$ is being considered \cite{uscensus:2021:discreteG}. However, this may come at the cost of requiring a fairly large $\eps$. Despite being only approximately differentially private, the Gaussian mechanism is fairly general and powerful and lends itself to analytical privacy formulations and sharp approximate, concentrated, or composition privacy bounds \cite{dwork2006calibrating,bun2016concentrated, balle18a, mironov2017renyi,dong2019gaussian,bun2018composable,Shahab_ISIT_2020:100rounds,CDP:dwork:2016, composition:theorems}.

Motivated by such considerations and tradeoffs and towards further improving the privacy performance in various applications of DP, we aim to answer this question: 

\emph{Is it possible to develop a mechanism for DP with sub-Gaussian tail which can provide better approximate and concentrated differential privacy performance than a comparable Gaussian mechanism?}

We answer this question in the affirmative and propose a new probability distribution for mechanism design, defined below, which is obtained by utilizing the normalized tails of two symmetric Gaussians around zero.

\begin{definition}(Offset-Symmetric Gaussian Tails Distribution)
Let $m, \sigma \in \R^+$ be given as input parameters. The offset-symmetric Gaussian tails (OSGT) probability density function (pdf) is constructed by using the respective positive and negative (normalized) tails of two Gaussian distributions $\N(-m,\sigma^2)$ and $\N(m,\sigma^2)$ as follows
\begin{align}\label{eq:osgt1}
  f(y)=\begin{cases}
        \frac{1}{S} \exp({-\frac{\left (y-m  \right )^{2}}{2\sigma ^{2}}}), & y\leq 0,\\
      \frac{1}{S} \exp({-\frac{\left (y+m  \right )^{2}}{2\sigma ^{2}}}), & y> 0,
      \end{cases}
\end{align}
where $S$ is the normalization constant to make the pdf  valid and is readily determined by considering the area under the two offset-symmetric Gaussian tails as $S = 2\sqrt{2\pi\sigma^2} Q({m}/{\sigma})$, where $Q(\cdot)$ is the Gaussian $Q$-function: $Q(x) = \frac{1}{\sqrt{2\pi}}\int_x^\infty e^{-{y^2}/{2}} dy.$\footnote{In this paper, we use both $\exp(x)$ and $e^x$ for the exponential function while taking into account aesthetic, space and readability issues.}
\end{definition}\label{def:osgt}

\begin{figure}[t]
\begin{tikzpicture}\label{osgtfig}
\begin{axis}[width=100mm, height=40mm,axis lines=middle,axis line style = thick,
		xtick={-\m, \m}, xticklabels={$\smash{-}m$, $m$},
		ymax=1.1,
		ytick=\empty, yticklabels=\empty,
		every tick/.style={black,semithick,major tick length = 3mm}]
	\addplot[name path=minterm,line width=1pt]
		expression[domain=-\w:\w,samples=191] {min(exp(-0.5*(x+\m)^2),exp(-0.5*(x-\m)^2))} ;
	\addplot[line width=1pt,dotted]
		expression[domain=-\w:\w,samples=191] {max(exp(-0.5*(x+\m)^2),exp(-0.5*(x-\m)^2))} ;
	\addplot +[mark=none,color=black,dotted] coordinates {(\m, 0) (\m, 1.05)} ;
	\addplot +[mark=none,color=black,dotted] coordinates {(-\m, 0) (-\m, 1.05)} ;
	\path[name path=axis] (axis cs:-\w,0) -- (axis cs:\w,0) ;
	\addplot[thick,fill=black!50!white,fill opacity=0.5]
		fill between[of=minterm and axis, soft clip={domain=-\w:\w}] ;
\end{axis}
\end{tikzpicture}
   \caption{The normalized tails of two Gaussians at $\pm m$ with variance $\sigma^2$ make the OSGT distribution.}
    \label{fig:osgt}
\end{figure}
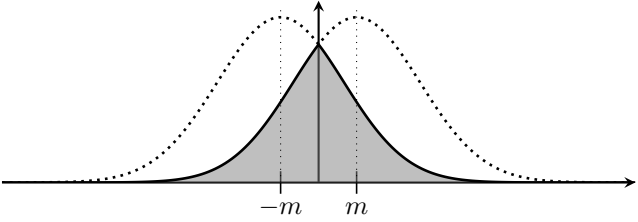

A schematic of the OSGT distribution is shown in Fig. \ref{fig:osgt}.

We say that a random variable $Y$ has the OSGT distribution, which is denoted as $Y \leftarrow \OSGT$, if its pdf follows \eqref{eq:osgt1}. Due to the symmetry of the OSGT pdf, $Y$ has zero mean. The cumulative distribution function (cdf) of the OSGT random variable $Y$ is given by
\begin{equation}
  F(y)\doteq\Prob[Y < y] = \begin{cases}
        \frac{Q\left(({m-y})/{\sigma}\right)}{2Q\left({m}/{\sigma}\right)}, & y\leq 0,\\
      1-\frac{Q\left(({m+y})/{\sigma}\right)}{2Q\left({m}/{\sigma}\right)}, & y> 0.
      \end{cases}
      \label{cdf}
\end{equation}
Its variance $V(m, \sigma^2)$, or $V$ for short, is a function of the input parameters $m ,\sigma^2$ and will be detailed later.  
\subsection{Related Work}\label{sec:related}
Differential privacy (DP) was proposed in the pioneering work of~\cite{dworkoriginal2006,dwork2006calibrating, dwork2006our}. See~\cite{dwork2014algorithmic} for fundamentals of DP. Our work focuses on mechanism design and privacy performance analysis. 

The first mechanism for DP was the Laplace mechanism~\cite{dworkoriginal2006,dwork2006calibrating, dwork2006our}. Nissim, et al.~\cite{nissim2007smooth} proposed to improve utility via scaling the Laplace distribution by the smooth sensitivity of the query function.~In \cite{mironov2012significance}, Mironov demonstrated a serious privacy vulnerability in `off-the-shelf' implementations of the Laplace mechanism on finite-precision computers and proposed a snapping mechanism to address this issue. A similar problem was also studied in~\cite{gazeau2016preserving}. In \cite{dwork2006our}, Dwork et al.~studied the binomial mechanism, which was improved by Agarwal et al.~in~\cite{agarwal2018cpsgd}. Works that have extensively studied the Gaussian mechanism include~\cite{bun2016concentrated, balle18a,dong2019gaussian}. The discrete versions of the Laplace and Gaussian mechanisms have respectively been studied in works such as~\cite{ghosh2012universally,NEURIPS2020_b53b3a3d}.

To our best knowledge, there has been little work on implementing a DP mechanism, whose tail is sub-Gaussian, but can provide better approximate and concentrated DP performance than a comparable Gaussian mechanism.~One such mechanism is proposed here, which is obtained by utilizing the normalized tails of two symmetric Gaussians around zero.~Thanks to its ``Gaussian origin", analytical derivations and exact discrete sampling of our mechanism is possible and its applications can come naturally.~Recently,~\cite{GenGaussLiu2019} proposed the generalized Gaussian mechanism for DP, which can provide lighter tails than the Gaussian for distribution orders $p > 2$. However, this mechanism does not lend itself to analytical privacy performance derivations or designs. In~\cite{ganesh_FORC_21}, the worst-case error performance of DP was bounded using generalized Gaussian mechanisms. We are not aware of any work on discrete generalized Gaussian mechanisms.

\subsection{Main Contributions}
The contributions of this paper are summarized as
follows:
\begin{itemize}
    \item We study the properties of the single-variate OSGT distribution analytically and also provide useful insights into its behavior as follows. We argue that the ratio ${m}/{\sigma}$ plays an important role in its behavior. When ${m}/{\sigma} \to 0$, the OSGT distribution tends to the Gaussian distribution $\N(0,\sigma^2)$, as expected. For intermediate values of ${m}/{\sigma}$, the pdf displays some sort of hybrid behavior between (scaled) Laplace and Gaussian distributions.~We detail in Proposition \ref{prop:var} the exact variance $V(m, \sigma^2)$. In Propositions \ref{prop:sufficient}-\ref{prop:4}, we show that the OSGT distribution remains $\sigma^2$-sub-Gaussian for sufficiently small ${m}/{\sigma}$,\footnote{Specifically for ${m}/{\sigma} \le Q^{-1}(0.25) \approx 0.6745$; see Proposition \ref{prop:sufficient}.} and generally provides a tradeoff in tail, in between Gaussian and Laplace tails (see Fig. \ref{Fig_tail} for an example). We extend the OSGT distribution to the independent and identically distributed (i.i.d.) multi-variate case for use in the $k$-dimensional OSGT mechanism design.

    \item Further, we exactly derive the $(\epsilon, \delta(\eps))$-differential privacy performance of the single-dimensional  OSGT mechanism. We provide several numerical examples (see Fig. \ref{fig:delta1:delta2}) where the OSGT mechanism ensures a better $\delta(\eps)$ privacy protection than the Gaussian mechanism for the same variance $V$ and at the same $\eps$. Therefore, the OSGT mechanism has the potential to be used in larger datasets and can provide an overall better privacy performance with a comparable mean-square error (variance).

    \item An alternative method to characterize the privacy performance of a mechanism is the R\'enyi divergence-based or concentrated differential privacy \cite{mironov2017renyi, bun2016concentrated}. We study the zero-concentrated differential privacy (zCDP) performance of the $k$-dimensional OSGT mechanism and derive a single-letter analytical upper bound on it. We show that the OSGT mechanism provides $(\zeta,\frac{\Delta_2^2}{\sigma^{2}})$-zCDP, where  $\Delta_2$ is the $2$-norm of the global query function sensitivity and $\zeta$ is a constant, where $\zeta \to 0$ for sufficiently large concentration orders $\alpha$ (see Theorem \ref{thm:zcdp}). We will show that for fixed $k, m, \Delta_2$ and $\sigma^{2}$, and for sufficiently large $\alpha$, the concentration of differential privacy for the OSGT mechanism is better than that of the Gaussian mechanism.

    \item Due to challenges in analytical computation of the exact $\delta(\eps)$ for the OSGT mechanism with $k > 1$ dimensions, we take an indirect approach to convert the R\'enyi divergence of the OSGT mechanism to an achievable bound on its $\delta(\eps)$. Using this result, we show that even for multi-dimensional queries, the OSGT mechanism can provide a better privacy performance, see Section \ref{sec:zcdpconvert}  for more details and Fig. \ref{fig:zcdp2delta} for an example.
    
\end{itemize}

This paper is organized as follows. In Section \ref{sec:prelim}, we review essential background on DP and various mechanisms. In Section \ref{secproperties}, we introduce the properties of the proposed OSGT distribution analytically and also provide useful insights into its behavior. In Section \ref{sec:privacy}, we present the proposed OSGT mechanism and analyze its privacy performance including its approximate DP for the single-dimensional query case, its zCDP performance for the multi-dimensional query case, and the conversion of the concentrated DP to the approximate DP for the multi-dimensional query case. We conclude the paper in Section \ref{sec:conclusion}.

\section{Preliminaries}\label{sec:prelim}
The following definitions are from~\cite{dwork2006calibrating, dwork2006our, bun2016concentrated, dwork2016concentrated, dwork2014algorithmic, balle18a}, which will be heavily used in this paper. 
\begin{definition}(Pure/Approximate Differential Privacy) A randomized mechanism $M: \X^n \to \Y$ is said to satisfy $(\epsilon, \del)$-differential privacy or $(\epsilon, \del)$-DP for short, if for all $x,x'\in \X^n$ differing on a single element and all events $E \subset \Y$, we have 
$$\Prob[M(x) \in E ]\leq e^\eps \Prob[M(x') \in E] +\del.$$
\end{definition}
Note that $(\epsilon, 0)$-DP is also known as  pure differential privacy and when $\del > 0$, $(\epsilon, \del)$-DP is known as approximate differential privacy. 

\begin{definition}(Zero-Concentrated Differential Privacy)\label{def:zcdp} A randomized mechanism $M: \X^n \to \Y$ is said to satisfy $(\zeta, \rho)$-zero-concentrated differential privacy or $(\zeta, \rho)$-zCDP for short, if for all $x,x'\in \X^n$ differing on a single element and all $\alpha \in (1, \infty)$, we have 
$$D_{\al}(M(x)\lVert M(x')) \leq \zeta + \rho\alpha,$$
where $D_{\al}(P\lVert Q) = \frac{1}{\al-1}\log\int_{y \in \Y} P(y)^\al Q(y)^{(1-\al)} dy$ is the R\'enyi divergence of order $\alpha$ of the distribution $P$ from the distribution $Q$. $\rho$-zCDP refers to $(0,\rho)$-zCDP.
\end{definition}
Note that pure $(\epsilon, 0)$-DP implies $\frac{1}{2}\eps^2$-zCDP \cite{bun2016concentrated}.

In this paper, we will focus on output perturbation DP mechanisms. Roughly speaking, an output perturbation mechanism operates on the output of a deterministic real-valued vector computation, called the query, and adds i.i.d. noise to each component of the query, which is drawn from a suitable multi-variate probability distribution. Specifically, the query function $q: \X^n \to \R^k$, where $q(x) = (q(x)_1, q(x)_2, \cdots, q(x)_k) \in \R^k$ computes the $k$-dimensional deterministic function of interest on dataset $x$. When the context is clear, we will drop the dependence of $q(x)$ on $x$ and simply write $q = (q_1, q_2, \cdots, q_k)$. Similarly, $q(x') = (q(x')_1, q(x')_2, \cdots, q(x')_k)$ operating on dataset $x'$ is simplified to just $q' = (q'_1, q'_2, \cdots, q'_k)$. In this paper, we may use the notation $x\sim x'$ to refer to any two neighboring datasets $x,x'\in\X^n$ differing on a single element.

\begin{definition}(Global Sensitivity)\label{def:global} A query function $q: \X^n \to \R^k$ is said to have global $p$-sensitivity $\Delta_p$ if $$\sup_{x,x'\in \X^n: x\sim x'}\norm{q(x)-q(x')}_p = \Delta_p,$$ where $\norm{\cdot}_p$ is the $p$-norm.
\end{definition}

\begin{definition}(Output Perturbation Mechanism) A randomized output perturbation mechanism $M: \X^n \to \R^k$ for the query function $q: \X^n \to \R^k$ is given by
$M(x) = q(x)+Y$, where $Y = (Y_1, \cdots, Y_k)$ and each $Y_i$ is a random variable drawn according to an i.i.d., but otherwise arbitrary probability distribution.
\end{definition}
We now review two important classes of output perturbation DP mechanisms in the literature.

\textbf{The Laplace mechanism \cite{dworkoriginal2006,dwork2006calibrating, dwork2006our}:} The Laplace mechanism $M: \X^n \to \R^k$ for the query function $q: \X^n \to \R^k$ is obtained as $M(x) = q(x)+Y$ where each noise component $Y_i$ in $Y = (Y_1, \cdots, Y_k)$ is drawn from a scalar i.i.d. Laplace distribution $\Lap(0, \lambda)$ at location $0$ and with the scale $\lambda$. 
For the Laplace mechanism, the pdf of $M(x)$ evaluated at $y \in \R^k$ is given as 
\begin{equation}\nonumber
    \begin{split}
    f_{M(x)}^{\Lap}(y) &= \prod_{i=1}^{k}\frac{1}{2\lambda}\exp\left(\frac{-|y-q_i|}{\lambda}\right)\\ &= \frac{1}{(2\lambda)^k}\exp\left(\frac{-\sum_{i=1}^k|y-q_i|}{\lambda}\right)\\ &= \frac{1}{(2\lambda)^k} \exp\left(\frac{-\norm{y-q}_1}{\lambda}\right).    
    \end{split}
\end{equation}
When the context is clear, we may drop the $\Lap$ superscript denoting the Laplace mechanism. Also, when the dependence of the density on $M(x)$ is clear from the location $q$, we may drop the subscript $M(x)$. For the 1-sensitivity of the query $\Delta_1 = \sup_{x,x':x\sim x'}\norm{q(x)-q(x')}_1$ and for $\lambda = {\Delta_1}/{\eps}$, the Laplace mechanism will achieve $(\epsilon, 0)$-DP.

\textbf{The Gaussian mechanism \cite{balle18a}:} The Gaussian mechanism $M: \X^n \to \R^k$ for the query function $q: \X^n \to \R^k$ is obtained when $M(x) = q(x)+Y$ and each noise component $Y_i$ in $Y = (Y_1, \cdots, Y_k)$ is drawn from a scalar i.i.d.  Gaussian distribution $\N(0, \sigma_g^2)$ at location 0 and with the scale parameter $\sigma_g$. For the Gaussian mechanism, the pdf of $M(x)$ evaluated at $y \in \R^k$ is given as
\begin{equation}\nonumber
    \begin{split}
     f_{M(x)}^{\N}(y) &= \prod_{i=1}^{k}\frac{1}{\sqrt{2\pi\sigma^2_g}}\exp\left(\frac{-(y-q_i)^2}{2\sigma^2_g}\right)\\ &=  \frac{1}{\left(\sqrt{2\pi\sigma^2_g}\right)^k} \exp\left(\frac{-\norm{y-q}_2^2}{2\sigma^2_g}\right).   
    \end{split}
\end{equation}
When the context is clear, we may drop the $\N$ superscript denoting the Gaussian mechanism. Also, when the dependence of the density on $M(x)$ is clear from the location $q$, we may drop the subscript $M(x)$.
For the global 2-sensitivity of the query $\Delta_2 = \sup_{x,x':x\sim x'}\norm{q(x)-q(x')}_2$, the mechanism achieves $\frac{\Delta_2^2}{2\sigma_g^2}$-zCDP. For example, if one sets $\sigma^2_g = {\Delta_2^2}/{\eps^2}$, then $\frac{1}{2}\eps^2$-zCDP will be achieved.

\begin{definition}(Privacy Loss Random Variable)\label{def:PL}
Let $M: \X^n \to \R^k$ be an output perturbation mechanism and $x, x'\in \X^n$ be two neighboring datasets differing on a single element. Let $f_{M(x)}$ be the probability density function of the random variable $M(x)$ and  $f_{M(x')}$ be the probability density function of the random variable $M(x')$. Define the privacy loss function of the mechanism $M$ on the pair of inputs $x,x'$ as
\begin{align}\label{eq:PLRV}
\ell_{M,x,x'}(y) = \log\left(\frac{f_{M(x)}(y)}{f_{M(x')}(y)}\right).
\end{align}
 Let $L = \ell_{M,x,x'}(M(x))$ be the transformation of the output random variable $M(x)$ by the function $\ell_{M,x,x'}$. 
\end{definition}
The privacy loss random variable is useful in determining $\del$ as a function of $\eps$ in the $(\epsilon, \del)$-DP framework as follows:
$$\delta(\eps) = \Prob[L \geq \eps] - e^{\eps} \Prob[L' \leq -\eps],$$
where $L' = \ell_{M,x',x}(M(x'))$ is the transformation of the output random variable $M(x')$ by the function $\ell_{M,x',x}$. 

For the Gaussian mechanism, this yields \cite{balle18a}
\begin{equation}\nonumber
    \begin{split}
     \delta^\N(\eps) &= \Prob[L \geq \eps] - e^{\eps} \Prob[L' \leq -\eps]\\
    &=\underset{Y\leftarrow \N(0,\sigma^2_g)}{\Prob}\left[Y > \frac{\sigma^2_g\eps}{\Delta_2}-\frac{\Delta_2}{2}\right]\\
    &\quad{}-e^{\eps}\underset{Y\leftarrow \N(0,\sigma^2_g)}{\Prob}\left[Y > \frac{\sigma^2_g\eps}{\Delta_2}+\frac{\Delta_2}{2}\right].   
    \end{split}
\end{equation}
The value of $\delta_g^{\N}(\eps)$ can be computed using the Gaussian $Q$-function. Depending on the relation of $\eps$ and $\frac{\Delta_2 ^2}{2\sigma^2_g}$, we can detail $\delta_g^{\N}(\eps)$ for the Gaussian mechanism as
\begin{align*}
\del^\N(\eps)=\begin{cases}
        Q\left(\frac{\sigma_g\eps}{\Delta_2}-\frac{\Delta_2}{2\sigma_g}\right)\!-\!e^\eps Q\left(\frac{\sigma_g\eps}{\Delta_2}+\frac{\Delta_2}{2\sigma_g}\right), & \hspace{-0.6em}\eps \geq \frac{\Delta_2 ^2}{2\sigma^2_g},\\
        1\!-\!Q\left(\frac{\Delta_2}{2\sigma_g}\!-\!\frac{\sigma_g\eps}{\Delta_2}\right)\!-\!e^\eps Q\left(\frac{\sigma_g\eps}{\Delta_2}+\frac{\Delta_2}{2\sigma_g}\right), & \hspace{-0.5em} \eps < \frac{\Delta_2 ^2}{2\sigma^2_g}.
      \end{cases}
\end{align*}

\section{Properties of the OSGT Distribution}\label{secproperties}
In this section, we will study the OSGT distribution. We will derive the variance of the OSGT-distributed random variable in a simple analytical form and compare its pdf and tail with those of the Gaussian and the Laplace distributions. For the ease of analysis, we will mainly focus on the one-dimensional or single-variate case. However, we will present a special i.i.d. multi-variate OSGT distribution, which will be used in Section \ref{sec:privacy} for the OSGT mechanism. 

\subsection{Single-variate Case}
Let us recall the definition of the single-variate OSGT probability density function from \eqref{eq:osgt1}:
\begin{equation}\label{eq:osgt2}
  f(y)=\begin{cases}
        \frac{1}{S} \exp\left({-\frac{\left (y-m  \right )^{2}}{2\sigma ^{2}}}\right), & y\leq 0,\\
      \frac{1}{S} \exp\left({-\frac{\left (y+m  \right )^{2}}{2\sigma ^{2}}}\right), & y> 0.
      \end{cases}
\end{equation}
We say a random variable is OSGT distributed and denote it by $Y \leftarrow \OSGT$ if its pdf follows \eqref{eq:osgt2}. Note that in this definition, $m, \sigma^2 \in \R^+$ are \emph{input parameters} to the distribution, but are not its actual location or scale. The actual mean of $Y$ is zero and its variance $V$ is a function of $m, \sigma^2$. That is why we denote the distribution by $\OSGT$ to signify this dependence.\footnote{When we compare the OSGT distribution with the Laplace or Gaussian distributions, we will compare them at the same actual variance. That is, we will set $\sigma^2_g = V(m,\sigma^2)$ for the Gaussian distribution and $2\lambda^2 = V(m,\sigma^2)$ for the Laplace distribution.} We will use $\mathcal{T}$ for \emph{tail} to emphasize that the distribution is built by using the respective positive and negative (normalized) tails of two Gaussian distributions $\N(-m,\sigma^2)$ and $\N(m,\sigma^2)$.

In some of our upcoming derivations, it will be useful to write the OSGT pdf as a single-case equation by taking advantage of the absolute function. To see this, let us write $f(y)$ in \eqref{eq:osgt2} as
\begin{align}
  f(y)=\begin{cases}
   \frac{1}{S}\exp\left({\frac{-m^{2}}{2\sigma ^{2}}}\right)\exp\left({\frac{-y^2+2my}{2\sigma ^{2}}}\right),  &y\leq 0,\\ \frac{1}{S}\exp\left({\frac{-m^{2}}{2\sigma ^{2}}}\right)\exp\left({\frac{-y^2-2my}{2\sigma ^{2}}}\right),  &y> 0,
        \end{cases}
\end{align}
from which we conclude
\begin{align}\label{eq:osgt1case}
  f(y)=
   \frac{1}{S'}\exp\left({-\frac{y^2}{2\sigma^{2}}}{-\frac{m|y|}{\sigma^{2}}}\right),
\end{align}
where $S' = 2\sqrt{2\pi\sigma^2} \exp({{m^{2}}/{(2\sigma ^{2})}})Q({m}/{\sigma}) = \exp({{m^{2}}/{(2\sigma ^{2}})}) S$.
The OSGT pdf at location $\mu$ is given as
\begin{align}\label{eq:multi}
  f_{\mu}(y)=
   \frac{1}{S'}\exp\left({-\frac{(y-\mu)^2}{2\sigma^{2}}}-\frac{m|y-\mu|}{\sigma^{2}}\right).
\end{align}
It is insightful to understand the general behavior of the (zero-mean) OSGT distribution for various $m, \sigma^2$ input parameters. Let us rewrite \eqref{eq:osgt1case} as 
\begin{align}\label{eq:intuition}
  f(y)=
   \frac{1}{S'}\exp\left(-\frac{|y|(|y|+2m)}{2\sigma^2}\right).
\end{align}
Whenever $|y|$ and $m$ are such that  $|y| \ll 2m$, we can we can ignore $|y|$ in the second term in the numerator of the exponential and approximate the OSGT pdf by  $f(y)\approx {1}/{S'}\exp(-m|y|/{\sigma^2})$, which is a scaled version of the Laplace pdf at location 0 and with scale $\lambda = \sigma^2/m$. Whenever $m/\sigma \ll 1$, the OSGT pdf  approximates a Gaussian pdf at location 0 and with scale $\sigma$. Therefore, it is intuitively understood that the OSGT distribution is some sort of \emph{hybrid} between Gaussian and Laplace distributions. In the remaining part of this section, we will make formal statements about the variance and tail of the OSGT distribution.

\subsubsection{Variance}
The following proposition provides the variance of an OSGT-distributed random variable. When used as noise in the output perturbation differential privacy mechanism in Section \ref{sec:privacy}, the variance of the OSGT mechanism provides a way to measure its utility. Specifically, the variance measures the mean square error of the perturbed output random variable $M(x)$ from the true deterministic query value $q(x)$. Generally speaking, it should be understood that the lower the variance of the noise, the higher the utility (or the lower the mean square error distortion) of the mechanism. 

\begin{proposition}\label{prop:var}
Let $Y \leftarrow \OSGT$ be a zero-mean OSGT distributed random variable. Its variance is given by
\begin{align}
\mathbb{E}[Y^2] = V(m, \sigma^2) = \sigma^2+m^2-\frac{m\sigma \exp\left(-\frac{m^2}{2\sigma^2}\right)}{\sqrt{2\pi}Q\left(\frac{m}{\sigma}\right)}< \sigma^2.\label{eqvar}
\end{align}
\end{proposition}
\begin{IEEEproof}
Note that due to the symmetry of the distribution, we have
\begin{equation}
    \begin{split}
     \mathbb{E}[Y^2] &= \frac{2}{S}\int_0^\infty y^2 \exp\left(-\frac{\left (y+m  \right )^{2}}{2\sigma ^{2}}\right) dy\\\label{eq5} &=
\frac{\exp({-\frac{m^{2}}{2\sigma^2}})}{\sqrt{2\pi}\sigma Q\left(\frac{m}{\sigma}\right)}\int_0^\infty y^2 \exp\left({-\frac{y^2+2my}{2\sigma ^{2}}}\right) dy.  
    \end{split}
\end{equation}

To simplify the integral, we utilize the readily available formula 3.462.1 in \cite{gradshteyn2007}, which is repeated here:
\begin{equation}
    \begin{split}
     &\int_{0}^{\infty}x^{v-1}\exp({-\beta x^{2}-\gamma x})dx\\\label{eq6}
     &=\left ( 2\beta \right )^{\frac{-v}{2}}\Gamma\left ( v \right )\exp\left({\frac{\gamma^{2}}{8\beta}}\right)D_{-v}\left ( \frac{\gamma}{\sqrt{2\beta}} \right ),   
    \end{split}
\end{equation}
where $\Gamma(\cdot)$ is the Gamma function and $D_{-v}(\cdot)$ is the parabolic cylinder function. For $D_{-v}(\cdot)$, formulae 9.247.1-3 and 9.254.1-2 in \cite{gradshteyn2007}) are repeated here in terms of the $Q$-function:
\begin{equation}
    D_{-1}\left ( z \right )=e^{z^{2}/{4}}\sqrt{2\pi}Q (z)
    \label{eq7},
\end{equation}
\begin{equation}
    D_{-2}(z)=e^{-z^{2}/{4}}-e^{{z^{2}}/{4}}\sqrt{2\pi}zQ (z)
    \label{eq8},
\end{equation}
\begin{equation}
  D_{p+1}\left ( z \right )-zD_{p}\left ( z \right )+pD_{p-1}\left ( z \right )=0
    \label{eq9}.
\end{equation}
 We set $v=3$, $\beta={1}/{(2 \sigma^{2})}$, $\gamma={m}/{\sigma^2}$, $z = {\gamma}/{\sqrt{2\beta}} ={m}/{\sigma}$, and use the formulae above at $p=-2$ to obtain
\begin{equation}\nonumber
    \begin{split}
     D_{-3}\left ( \frac{m}{\sigma} \right )=\frac{1}{2}&\biggl ( \sqrt{2\pi}\exp\left({\frac{m^2}{4\sigma^{2}}}\right) Q\left ( \frac{m}{\sigma} \right ) \left ( 1+\frac{m^{2}}{\sigma^{2}}\right )\\
     &\quad{}-\frac{m}{\sigma}\exp\left({-\frac{m^{2}}{4\sigma^{2}}}
     \right)\biggr ).   
    \end{split}
\end{equation}
Therefore, using \eqref{eq6}, we have
\begin{equation}\nonumber
    \begin{split}
     &\int_0^\infty y^2 \exp\left(-\frac{\left (y+m  \right )^{2}}{2\sigma ^{2}}\right) dy\\ 
     &= \sqrt{2\pi}\exp\left({\frac{m^{2}}{2\sigma^{2}}}\right)\sigma^3Q\left ( \frac{m}{\sigma} \right )\left ( 1+\frac{m^{2}}{\sigma^{2}}\right )-m\sigma^2.   
    \end{split}
\end{equation}
Using this equation in \eqref{eq5} gives the variance in \eqref{eqvar}. 
Now to show $V(m, \sigma^2) < \sigma^2$, it suffices to show $$\frac{m\sigma \exp\left({-{m^2}/{2\sigma^2}}\right)}{\sqrt{2\pi}Q\left({m}/{\sigma}\right)} > m^2,$$
which can easily be  verified since $Q(z) < \exp(-{z^2}/{2})/{(z\sqrt{2\pi})}$, $z > 0$, evaluated at $z = {m}/{\sigma}$.
\end{IEEEproof}

\begin{figure}
         \centering
         \includegraphics[width=8.8cm]{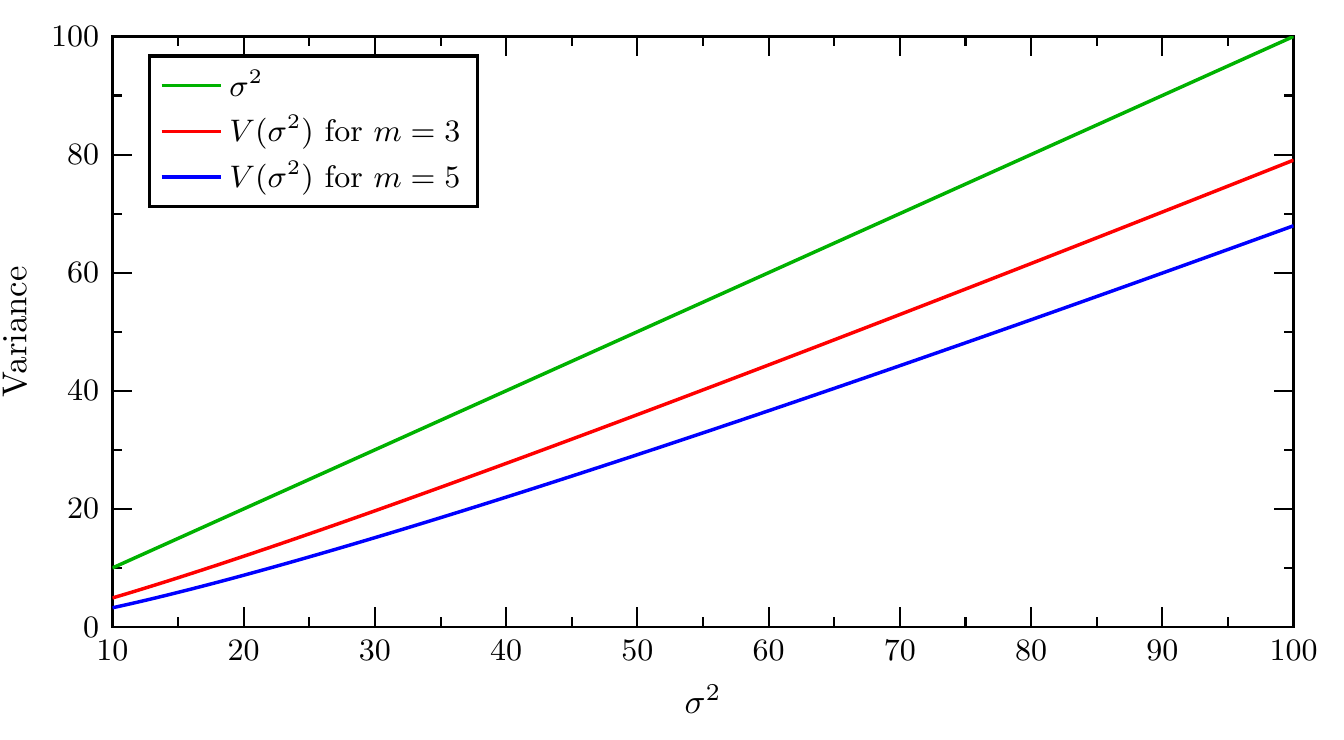}
     \caption{ Comparison of the input parameter $\sigma^2$ with the variance $V(m, \sigma^2)$ for the OSGT-distributed random variable $Y$ for two different values of $m$. It is clear that $V(m, \sigma^2)$ can be significantly smaller than $\sigma^2$ and the gap depends on both $m$ and $\sigma^2$. For a fixed $\sigma^2$, the gap widens as $m$ increases.}
    \label{Fig_var}
 \end{figure}

To summarize, the variance of an OSGT distributed random variable with input parameters $m$ and $\sigma^2$ is strictly less than $\sigma^2$ and our numerical results show the gap between $V(m, \sigma^2)$ and $\sigma^2$ depends on both $m$ and $\sigma^2$. See Fig. \ref{Fig_var}.

The fact that $V(m,\sigma^2) < \sigma^2$ has important positive implications for the privacy performance of the OSGT mechanism. The high-level reason is as follows. We will see in Section \ref{sec:privacy} that the  privacy performance of the OSGT mechanism will be a function of the input parameters $m$ and $\sigma^2$. When we compare the privacy  performance of the OSGT mechanism with that of the Gaussian mechanism, we will compare them at the same actual variance $\sigma^2_g = V(m, \sigma^2) < \sigma^2$ for a fair mean-square error comparison. We will see that the lower variance  of the Gaussian mechanism than the design parameter will translate into a worse privacy performance than the OSGT mechanism for many parameter regimes of interest. For more details, see Section \ref{sec:privacy}. 

\subsubsection{Tail Behavior}
Now, we turn our attention to the tail behavior of the OSGT distribution as another proxy for utility: a lighter tail will reduce the chance of outliers. This is an important consideration in many practical scenarios and was the main motivation for switching to the (discrete) Gaussian mechanism in the 2020 US Census release \cite{uscensus:2021:discreteG}.
\begin{proposition}\label{prop:sufficient} The OSGT distribution $\OSGT$ is $\sigma^2$-sub-Gaussian if ${m}/{\sigma} \leq Q^{-1}(0.25) \approx 0.6745$.
\end{proposition}
\begin{IEEEproof}
Let $y > 0$ be given.  Note that $m > 0$, $Q$-function is a decreasing function and $Q(y) \leq \frac{1}{2}\exp(-y^2/2)$ for $y >0$. Using \eqref{cdf}, we write
\begin{equation}\nonumber
    \begin{split}
      \underset{Y\leftarrow \OSGT}{\Prob}\left[Y\geq y\right] &= \frac{Q\left({(m+y)}/{\sigma}\right)}{2Q\left({m}/{\sigma}\right)} < \frac{Q\left({y}/{\sigma}\right)}{2Q\left({m}/{\sigma}\right)}\\ 
      &\leq \frac{\exp({-{y^2}/{2\sigma^2}})}{4Q\left({m}/{\sigma}\right)} \leq \exp\left({-\frac{y^2}{2\sigma^2}}\right), 
    \end{split}
\end{equation}
where the last inequality follows from the assumption ${m}/{\sigma} \leq Q^{-1}(0.25) \approx 0.6745$ or $4Q\left({m}/{\sigma}\right) \geq 1$. The last inequality establishes that $Y$ is $\sigma^2$-sub-Gaussian.

\end{IEEEproof}

We now provide two direct ways to compare the tails of two distributions, while taking into account their variance. Let us call $\bar{F}(y)=1-F(y)$ the survival function of the random variable $Y$ with cdf $F(\cdot)$.
\begin{definition}
 If $Y_{1}$ is heavier-tailed than $Y_{2}$, there exists $y_0$ where for all $y>y_{0}$,  $\bar{F}_{1}\left ( y \right )>\bar{F}_{2}\left ( y \right )$. 

\begin{definition}\label{def:tail2}
If $\lim_{y\rightarrow \infty}\frac{\bar{F}_{1}\left ( y \right )}{\bar{F}_{2}\left ( y \right )}=\infty$, then $Y_{1}$ has a heavier tail than $Y_{2}$. Equivalently, if $Y_{1}$ and $Y_{2}$ have respective probability density functions $f_{1}$ and $f_{2}$ and $\lim_{y\rightarrow \infty}\frac{f_{1}\left ( y \right )}{f_{2}\left ( y \right )}=\infty$, then $Y_{1}$ has a heavier tail than $Y_{2}$.
\end{definition}

\begin{remark}
The equivalence between the limits of the Survival functions and the probability density functions in Definition \ref{def:tail2} may not be true all the time. In Definition \ref{def:tail2}, the ratio of the survival functions tends to $\frac{0}{0}$ and we can use the L'Hopital's rule. We will specify the conditions for this to be valid: In Proposition \ref{prop:tailrel}, the survival function of the OSGT random variable is $\bar{F}_1(y)=\frac{Q\left(\frac{m+y}{\sigma}\right)}{2Q\left(\frac{m}{\sigma}\right)}$, which is differentiable and non-zero on the interval $I=[0,\infty)$. Likewise, the survival function of $\mathcal{N}(0,V={\sigma}_g^2)$ is $\bar{F}_2(y)=Q(\frac{y}{{\sigma}_g})$, which is also differentiable and non-zero on the same interval. Hence, we can use the L'Hopital's rule to find the limit of the ratio of survival functions using their respective pdfs. The same is true for the ratio of survival functions of the OSGT and the Laplace distributions used in Proposition \ref{prop:4}.
\end{remark}

\begin{proposition}\label{prop:tailrel}
The OSGT distribution has a heavier tail according to Definition \ref{def:tail2} than the normal distribution at the same variance $\sigma^2_g = V(m, \sigma^2)$.
\end{proposition}
\begin{IEEEproof}
See Appendix \ref{app_Gaustail}.
\end{IEEEproof}
\end{definition}
The proposition below formalizes that the tail of the Laplace distribution is heavier than that of the OSGT distribution for any variance $V(m,\sigma^2)$.

\begin{proposition}\label{prop:4}
The Laplace distribution has a heavier tail than the OSGT distribution according to Definition \ref{def:tail2}.
\end{proposition}

\begin{IEEEproof}
See Appendix \ref{app_Laptail}.
\end{IEEEproof}

 \begin{figure}
     \begin{subfigure}[b]{0.4\textwidth}
         \centering
         \includegraphics[width=8.8cm]{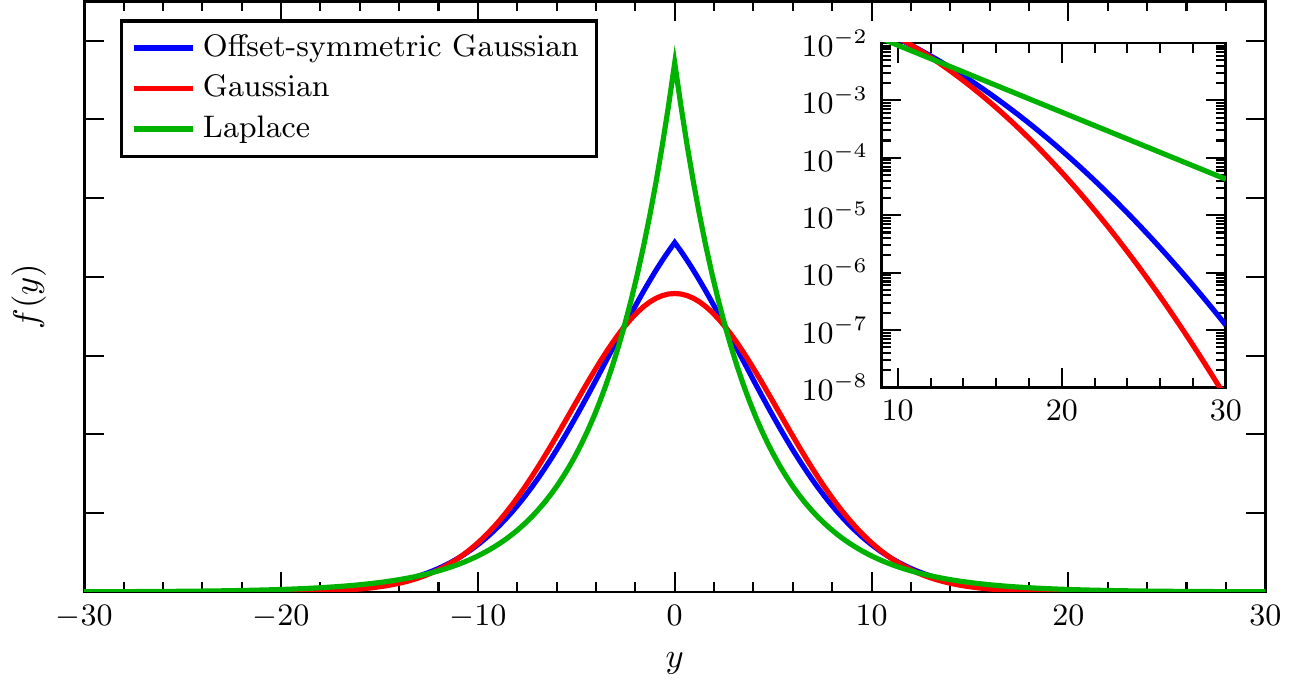}
         \caption{Probability density functions.}
         \label{fig:y equals x}
         \vspace{5mm}
     \end{subfigure}

     \begin{subfigure}[b]{0.4\textwidth}
         \centering
         \includegraphics[width=8.8cm]{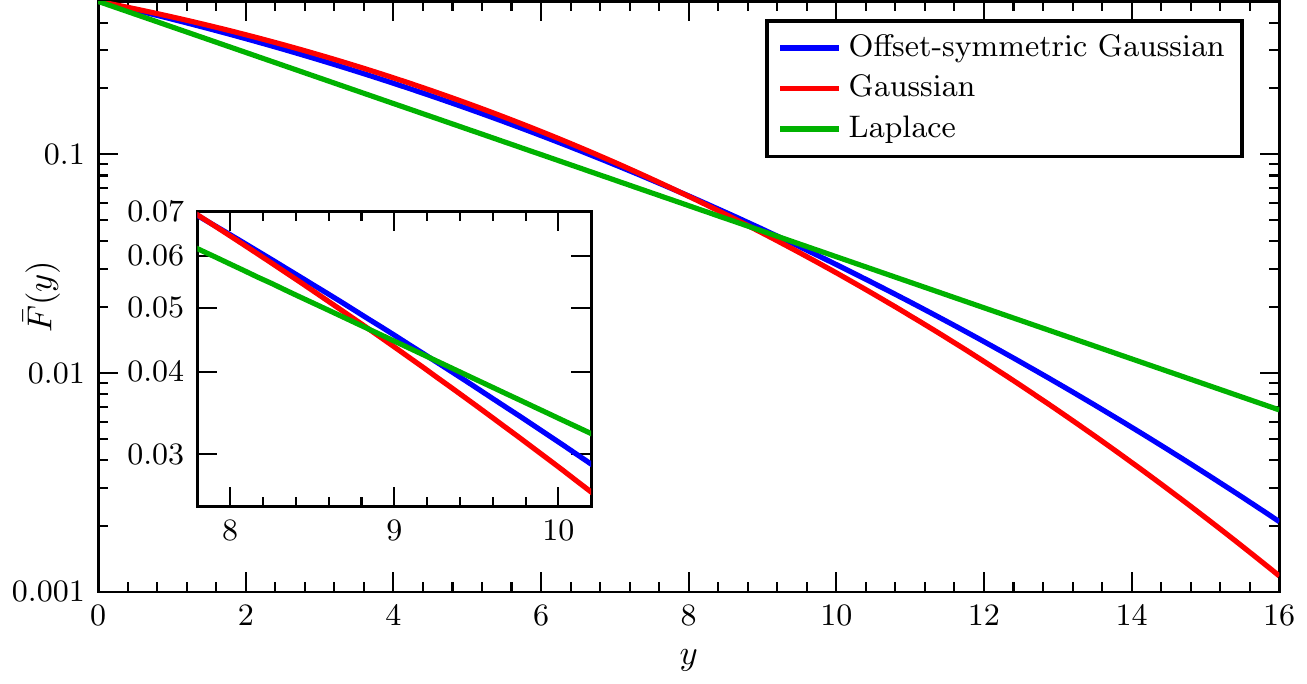}
         \caption{Survival functions.}
         \label{fig:three sin x}
     \end{subfigure}
     \caption{Comparison of the pdfs and their tails of the OSGT, Gaussian and Laplace distributions at the same effective variance $V(m, \sigma^2)$. We set $\sigma^{2}=40$ and $m=3$, which results in $V(m, \sigma^2) \approx 27.7047$ according to \eqref{eqvar}. It can be verified that ${m}/{\sigma} \approx 0.4743 \leq Q^{-1}(0.25) \approx 0.6745$, which means that the OSGT distribution is $\sigma^{2}$-sub-Gaussian. (a) the probability density function (b) the survival function for the corresponding distributions.}
    \label{Fig_tail}
 \end{figure}

We conclude this subsection by presenting in Fig. \ref{Fig_tail} a numerical comparison of the pdf and survival function of the OSGT, Gaussian and Laplace distributions. They are compared at the same effective variance for OSGT input parameters $m=3$ and $\sigma^2 = 40$, resulting in $V \approx 27.7047$. Our intuitive reasoning that the OSGT distribution is some sort of hybrid between and Gaussian and Laplace  distributions can be observed from Fig. \ref{Fig_tail}(\subref{fig:y equals x}). It can be seen from Fig. \ref{Fig_tail}(\subref{fig:three sin x}) that the tail (or the survival function) of the OSGT distribution lies between that of the Gaussian and Laplace distributions for large enough $y$.

\subsection{Multi-variate Case}
We consider a special extension of the scalar zero-mean OSGT random variable to the $k$-dimensional multi-variate case, which will be used in the OSGT mechanism design. More specifically, let $Y = (Y_1, \cdots, Y_k)$, where each $Y_i \leftarrow \OSGT$, $i \in [k]$, is i.i.d. distributed with zero mean and has the same input parameters $m, \sigma^2$. Overall, the multi-variate pdf is given by
\begin{align*}
 f(y_1, y_2, \cdots, y_k) &= \prod_{i=1}^k f(y_i) \\
 &= \prod_{i=1}^k\frac{1}{S'}\exp\left({-\frac{y_i^2}{2\sigma^{2}}}{-\frac{m|y_i|}{\sigma^{2}}}\right)\\&=\frac{1}{S'^k}\exp\left({-\frac{\sum_{i=1}^k y_i^2}{2\sigma^{2}}}{-\frac{m\sum_{i=1}^k|y_i|}{\sigma^{2}}}\right)\\
 &=\frac{1}{S'^k}\exp\left({-\frac{\norm{y}_2^2}{2\sigma^{2}}}{-\frac{m\norm{y}_1}{\sigma^{2}}}\right).
\end{align*}
We may denote such OSGT random vector by $Y \leftarrow \OSGTK$, where $I_k$ is the $k$-dimensional identity matrix.

It is interesting that the multi-variate OSGT pdf has a combination of $2$-norm and $1$-norm in its pdf. This combination can make some of the privacy analysis of the OSGT mechanism more challenging or at least, more involved. Nevertheless, using similar intuitive arguments as those provided around \eqref{eq:intuition}, we can argue that when all terms $|y_i|$ are much smaller than $2m$, the OSGT pdf can be approximated by the scaled pdf of a multi-variate i.i.d. Laplace. For $m/\sigma \ll 1$, the multi-variate OSGT pdf approximates the multi-variate Gaussian pdf $\N(0, \sigma^2I_k)$.
  
\section{Proposed OSGT Mechanism and Its Privacy Performance}\label{sec:privacy}

\begin{definition}\label{def:osgt:mech} The OSGT mechanism $M: \X^n \to \R^k$ for the query function $q: \X^n \to \R^k$ is obtained as $M(x) = q(x)+Y$ where each noise component $Y_i$ in $Y = (Y_1, \cdots, Y_k)$ is drawn from an i.i.d. scalar OSGT distribution $\OSGT$. For the OSGT mechanism, the pdf of $M(x)$ evaluated at $y \in \R^k$ is given as 
\begin{align*}
        f_{M(x)}^{\mathcal T}(y) 
 &= \prod_{i=1}^k\frac{1}{S'}\exp\left({-\frac{(y_i-q_i)^2}{2\sigma^{2}}}{-\frac{m|y_i-q_i|}{\sigma^{2}}}\right)\\&=
\frac{1}{S'^k}\exp\left({-\frac{\norm{y-q}_2^2}{2\sigma^{2}}}{-\frac{m\norm{y-q}_1}{\sigma^{2}}}\right).
\end{align*}

When the context is clear, we may drop the $\mathcal T$ superscript denoting the OSGT mechanism. Also, when the dependence of the density on $M(x)$ is clear from the location vector $q$, we may drop the subscript $M(x)$.   
\end{definition}

\subsection{Approximate Differential Privacy Performance}\label{sec:app:privacy}
In this section, we will focus on the single-dimensional query case where $k=1$ and derive the $(\epsilon, \delta)$ performance of the OSGT mechanism in \emph{exact} analytical form. That is, $M(x) = q(x)+Y$, where $q(x) \in \R$ and $Y \leftarrow \OSGT$ is a single OSGT random variable. Define $\Delta = \sup_{x,x'\in \X^n:x\sim x'}{|q(x')-q(x)|}$, which is the one-dimensional global query sensitivity. 

\begin{theorem}\label{thm:delta}
The single-dimensional OSGT mechanism $M(x) = q(x)+Y$,  $Y \leftarrow \OSGT$, achieves $(\epsilon, \del(\eps))$-DP, where $\del(\eps)$ is given by
\begin{equation}\label{eq:delta}
  \delta^{\mathcal{T}}(\eps)=
  \begin{cases}
   \begin{aligned}
      &1\!-\!\frac{1}{2Q\left({m}/{\sigma}\right)}\left(Q\left({1}/{2b}\!-\!b\eps\right)\!+\!e^\eps Q\left({1}/{2b}\!+\!b\eps\right)\right),
      \end{aligned}\\
\hspace{10em}\text{if} &\hspace{-11em}{\sigma^2\eps}/{\Delta} \leq {\Delta}/{2}+m,\\
   \begin{aligned}
        &\frac{1}{2Q\left({m}/{\sigma}\right)}\left(Q\left(a\eps\!-\!{1}/{2a}\right)\!-\!e^\eps Q\left(a\eps\!+\!{1}/{2a}\right)\right),
   \end{aligned}\\
        \hspace{10em} \text{if} & \hspace{-11em}{\sigma^2\eps}/{\Delta} > {\Delta}/{2}+m,
  \end{cases}
\end{equation}
where $a \doteq {\sigma}/{\Delta}$ and $b \doteq {\sigma}/{(2m+\Delta)}$.
\end{theorem}
\begin{IEEEproof}
See Appendix \ref{app_delta1D}.
\end{IEEEproof}
Here, we outline the ideas in the proof. In \cite{balle18a}, it was proved that the privacy loss random variable $L = \ell_{M,x,x'}$ for the $k$-dimensional Gaussian mechanism is itself a single-variate Gaussian. However, due to the more complex pdf of the OSGT distribution, which involves two cases (or an absolute value function), deriving the pdf of the privacy loss random variable $L = \ell_{M,x,x'}$ for the OSGT mechanism in analytical form, even for $k=1$, seems complicated. To see this, note that 
\begin{equation}\nonumber
    \begin{split}
     \log\left(\frac{f_{M(x)}(y)}{f_{M(x')}(y)}\right)&= \frac{(y-q')^2}{2\sigma^2} - \frac{(y-q)^2}{2\sigma^2}\\
     &\quad{}+ \frac{m|y-q'|}{\sigma^2}-\frac{m|y-q|}{\sigma^2}.
    \end{split}
\end{equation}
Instead, we take a first-principle approach to the derivation of $\delta$, which is inspired by the work \cite{balle18a}. First, for any two given mechanism random variables $M(x)$ and $M(x')$, located at $q=q(x)$ and $q'=q(x')$ with $\Delta_{q,q'}\doteq |q'-q|$ and midpoint $\bar{q} \doteq (q+q')/2$, we will find the worst-case set $E^*$
$$E^* = \{y \in \R: \log(f_{M(x)}(y)/f_{M(x')}(y)) \geq \eps \},$$
where the privacy loss function $\ell_{M,x,x'}$ in \eqref{eq:PLRV} is greater than $\eps$. We will prove that the set $E^*$ can be one of two different cases, detailed below, depending on the relation between $\eps$, $\Delta_{q,q'}$, $m$, and $\sigma^2$:
\begin{align*}
E^* = \begin{cases}
\left\{y: y \leq \bar{q}-\frac{\eps\sigma^2}{\Delta_{q,q'}+2m}\right\}, \quad & \frac{\sigma^2\eps}{\Delta_{q,q'}} \leq \frac{\Delta_{q,q'}}{2}+m, \\\\
\left\{y: y \leq \bar{q}+m-\frac{\eps\sigma^2}{\Delta_{q,q'}}\right\}, \quad & \frac{\sigma^2\eps}{\Delta_{q,q'}} >\frac{\Delta_{q,q'}}{2}+m.
\end{cases}
\end{align*}
Based on this set and assuming fixed $\eps$, $m$, and $\sigma^2$, we compute $\delta(\Delta_{q,q'})$ as the solution to the following integral
$$\delta(\Delta_{q,q'}) = \int_{E^*} (f_{M(x)}(y)-e^\eps f_{M(x')}(y))dy,$$
which as we will see in the detailed proof in Appendix \ref{app_delta1D} can be computed using the cdf of the OSGT distribution with appropriate location shifts. Finally, we will prove that when the parameters $\eps$, $m$, and $\sigma^2$ are fixed, $\delta(\Delta_{q,q'})$ is an increasing continuous function of $\Delta_{q,q'}$ and hence, it will assume its worst-case value at $\Delta = \sup_{x,x'\in \X^n:x\sim x'}{|q(x')-q(x)|}$, as given in \eqref{eq:delta}. We remark that as expected, at the boundary value $\eps = \frac{\Delta^2+2m\Delta}{2\sigma^2}$, $\del^{\mathcal T}(\eps)$ given in \eqref{eq:delta} is continuous:
$$\del^{\mathcal T}\left(\frac{\Delta^2+2m\Delta}{2\sigma^2}\right) = \frac{1}{2}-\frac{e^\eps Q\left(\frac{m+\Delta}{\sigma}\right)}{2Q\left(\frac{m}{\sigma}\right)}.$$
For more details, see Appendix \ref{app_delta1D}.

\begin{figure}
     \begin{subfigure}[b]{0.5\textwidth}
         \centering
         \includegraphics[width=8.8cm]{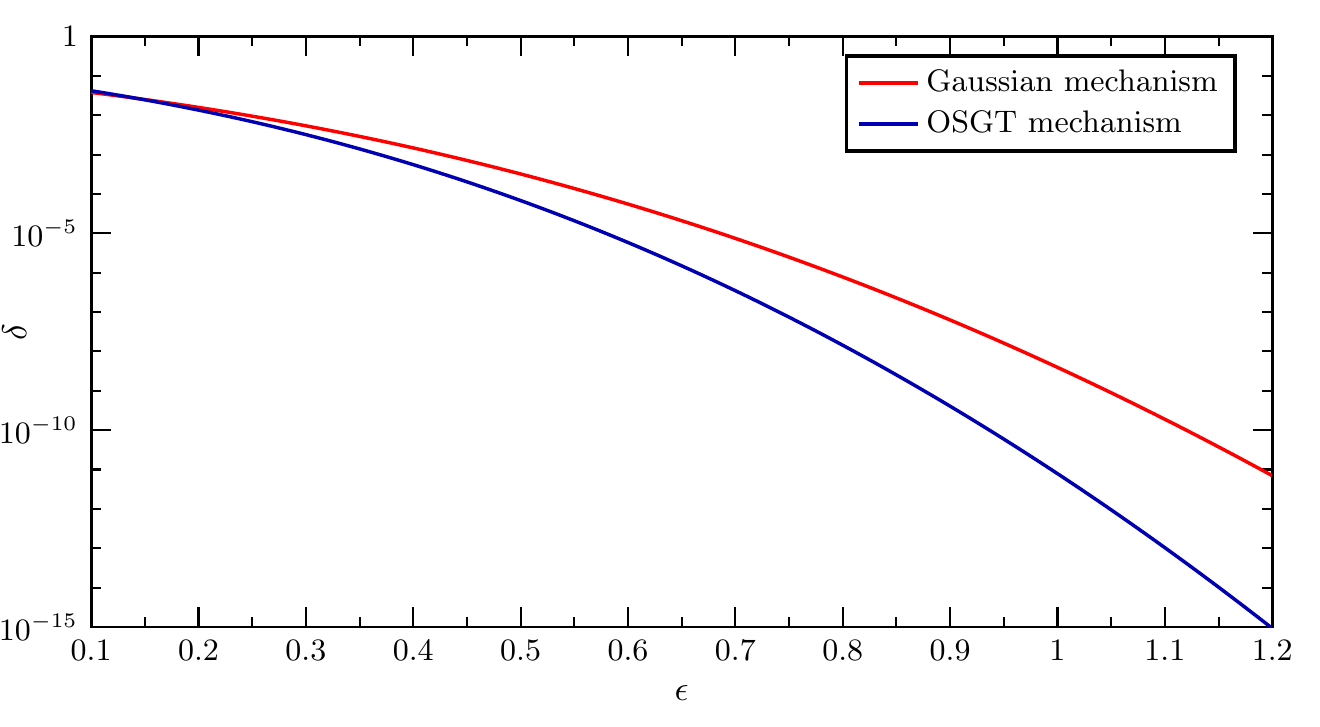}
         \caption{$m = 3$ and $\sigma^2 = 40$.}
         \label{fig:delta1}
     \end{subfigure}
     \hspace{-1em}
     \begin{subfigure}[b]{0.5\textwidth}
         \centering
         \includegraphics[width=8.8cm]{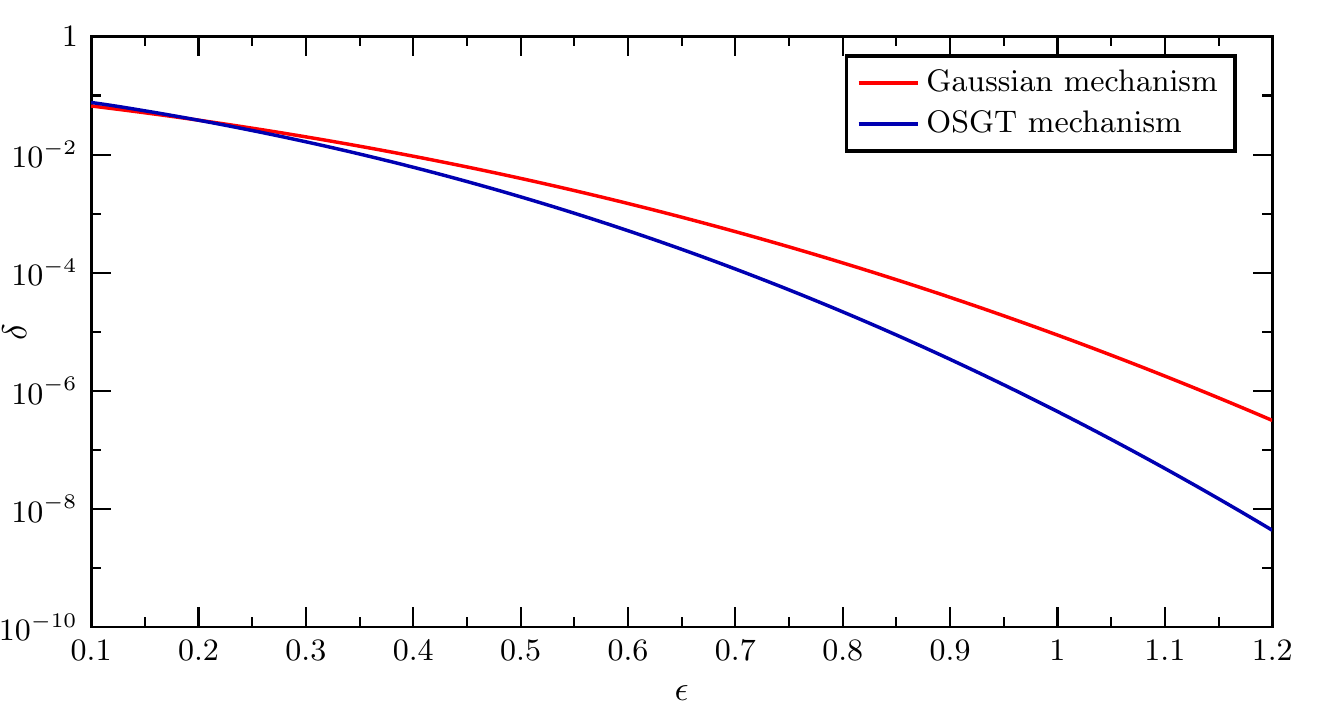}
         \caption{$m=2$ and $\sigma^2 = 20$.}
         \label{fig:delta2}
     \end{subfigure}
     \caption{ Comparing the approximate DP performance of $\OSGT$ and $\N(0,\sigma^2_g)$ mechanisms for $\sigma^2_g = V(m, \sigma^2)$, $\Delta = 1$ and two different $(m,\sigma^2)$ input tuples. It can be seen that $\delta(\eps)$ is generally lower for the OSGT mechanism than that for the Gaussian mechanism in the privacy regimes of practical interest where $\delta(\eps)$ must be very small.}
    \label{fig:delta1:delta2}
 \end{figure}

At $\sigma^2 = 40$, $m = 3$ and $\Delta = 1$, Fig. \ref{fig:delta1} compares the $(\epsilon, \delta(\eps))$ performance of the OSGT and Gaussian mechanisms at the same variance $V \approx 27.7047$ (note $V < \sigma^2$). For achieving $\del = 10^{-10}$, the OSGT mechanism requires $\eps \approx 0.94$, whereas the Gaussian mechanism requires $\eps \approx 1.12$. Alternatively, at equal $\eps = 1$, $\delta(\eps) \approx 7.8 \times 10^{-12}$ is achieved for the OSGT mechanism, whereas $\delta(\eps) \approx 3.9 \times 10^{-9}$ for the Gaussian mechanism. This demonstrates that $\delta(\eps)$ can be significantly worse for the Gaussian mechanism than that of the OSGT mechanism. Another set of results is shown in Fig. \ref{fig:delta2} for $m=2$ and $\sigma^2 = 20$. Note that $m/\sigma \approx 0.4472$ in this case and hence, the OSGT distribution is still $\sigma^2$-sub-Gaussian according to the sufficient condition of Lemma \ref{prop:sufficient}. We can observe again that the $(\epsilon, \delta)$ privacy performance of the OSGT mechanism is better than that of the Gaussian mechanism for the range of $\eps$ where $\delta(\eps)$ must be small.

In summary, the results in Figs. \ref{fig:delta1} and \ref{fig:delta2} argue favorably for the potential of the OSGT mechanism to be used in a wide range of privacy regimes for the release of large datasets. 

The detailed proof for the characterization of the exact analytical $\delta(\eps)$ for the single-dimensional OSGT mechanism highlights the difficulty in characterizing it for the multi-variate case with $k > 1$. Based on the good performance of the exact $\delta(\eps)$ for $k=1$ as numerically demonstrated above, we conjecture that for $k > 1$, at the same co-variance matrix $V(m,\sigma^2) I_k$, the OSGT mechanism may achieve smaller $\delta(\eps)$ than the Gaussian mechanism. This conjecture is supported by our results in the next two subsections. In subsection \ref{sec:zcdp}, we show the zCDP performance of the multi-dimensional OSGT mechanism is better than that of the Gaussian mechanism for a wide range of parameters. Then in Subsection \ref{sec:zcdpconvert}, we convert the concentrated DP (R\'enyi divergence) measure to upper bounds on $\delta(\eps)$ for $k$-dimensional OSGT and Gaussian mechanisms through applying the conversion method in \cite{NEURIPS2020_b53b3a3d}. This will  again show the superior performance of the OSGT mechanism for many privacy regimes.
 
 \subsection{Zero-Concentrated Differential Privacy Performance}\label{sec:zcdp}
\begin{theorem}
Consider the $k$-dimensional OSGT mechanism, where $M(x) = q(x)+Y$ and each noise component $Y_i$ in $Y = (Y_1, \cdots, Y_k)$ is drawn from an i.i.d. scalar OSGT distribution $\OSGT$. This mechanism achieves $(\zeta, \rho)$-zCDP where  
\begin{align}\label{eq:zcdp:theorem}
D_{\al}(M(x)\lVert M(x')) &\leq \zeta + \alpha\rho, \\
\rho &= \frac{\Delta_2^2}{2\sigma^2},\\
\zeta &= \frac{k}{\alpha-1}\log\left(\frac{1-Q(m/\sigma)}{Q(m/\sigma)}\right).
\end{align}
\label{thm:zcdp}
\end{theorem}

\begin{IEEEproof}
See Appendix \ref{app_zcdp}.
\end{IEEEproof}
We discuss some useful insights for interpreting the result in Theorem \ref{thm:zcdp}. Note that for fixed $k$, $m$, and $\sigma$, we have $\zeta \to 0$ as $\alpha\to \infty$. Therefore, the  zCDP performance of the OSGT mechanism approaches $\rho$-zCDP for sufficiently large $\alpha$. Also, note that for $m/\sigma \to 0$, $\frac{1-Q(m/\sigma)}{Q(m/\sigma)} \to 1$, and $\zeta \to 0$ for any $\alpha$ or $k$. Hence, for $m/\sigma \to 0$ we will recover the zCDP results for the Gaussian mechanism $\N(0, \sigma^2)$ as expected. Finally and most importantly, note that if we want to compare the zCDP performance of the OSGT and Gaussian mechanisms at the same actual variance (as a fair indicator of their mean square error), we will need to set $\sigma^2_g = V(m, \sigma^2)$ for the Gaussian mechanism, where $V(m, \sigma^2)$ was derived in \eqref{eqvar} and proven to be $V(m, \sigma^2)< \sigma^2$. Therefore, for the same 2-query sensitivity $\Delta_2$, we will have 
\begin{align}
\rho^{\mathcal T} &= \frac{\Delta_2^2}{2\sigma^2} < \rho^{\N} = \frac{\Delta_2^2}{2\sigma^2_g},
\end{align}
where superscripts $\mathcal T$ and $\N$ specify the OSGT and Gaussian mechanism, respectively. Therefore, for sufficiently large $\alpha$ where we can ignore the effect of $\zeta$, the zCDP performance of the OSGT mechanism is provably superior to that of the Gaussian mechanism at the same variance (or i.i.d. covariance matrix). 

\begin{figure}
         \centering
         \includegraphics[width=8.8cm]{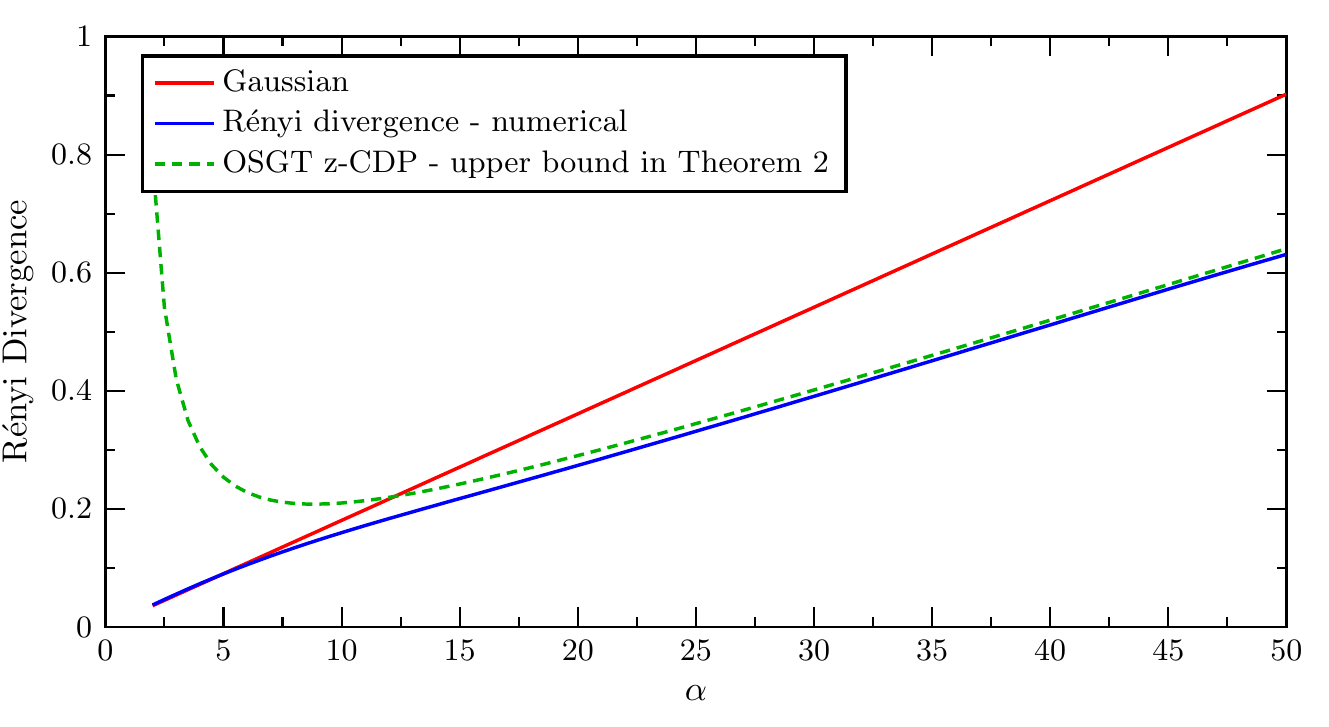}
     \caption{Comparison of the zCDP of Gaussian and OSGT mechanisms for $m = 3$, $\sigma^2 = 40$.} \label{fig:zcdp}
 \end{figure}

At $k=1$, $\sigma^2 = 40$, $m = 3$ and $\Delta = 1$, Fig. \ref{fig:zcdp} compares the differential privacy concentration of the OSGT and Gaussian mechanisms. Blue curve: we have numerically and directly computed the R\'enyi divergence of various orders $\alpha$. That is, the integral on the RHS of \eqref{eq:renyi} in Appendix \ref{app_zcdp} was numerically evaluated; see also Appendix \ref{app_renyi}. The dashed green curve shows the zCDP, which is the upper bound on the R\'enyi divergence. That is, the dashed green curve shows the RHS of \eqref{eq:zcdp:theorem}. The red curve is the $(0, \rho)$-zCDP of the Gaussian mechanism at the same variance $\sigma_g^2 = V(m, \sigma^2) \approx 27.7047$. The shape of the RHS of \eqref{eq:zcdp:theorem} for $\alpha < 10$ seems to be an artefact of our proof technique, which results in $\zeta$ showing large values for $\alpha \approx 1$, whereas the numerical evaluation of the RHS of \eqref{eq:renyi} scales well even at small $\alpha$. Overall, the superior zCDP performance of the OSGT mechanism is both analytically and numerically confirmed.

 \subsection{Conversion of Concentrated DP to Approximate DP}\label{sec:zcdpconvert}
 
As mentioned in Subsection \ref{sec:app:privacy}, analytical computation of the exact $\delta(\eps)$ for the OSGT mechanism with $k > 1$ dimensions in the query space seems challenging. Instead, we take an indirect approach to convert the concentrated DP performance of the OSGT mechanism to a bound on its $\delta(\eps)$. More specifically, by Proposition 12 in \cite{NEURIPS2020_b53b3a3d}, let $\alpha \in (1, \infty)$ and $\eps > 0$. If the R\'enyi differential privacy \cite{mironov2017renyi} satisfies
$$D_{\al}(M(x)\lVert M(x')) \leq \tau,$$
for all $x,x'\in \X^n$ differing on a single element, then
\begin{equation}\label{eq:delta:bound}
 \delta(\eps) = \frac{\exp((\alpha-1)(\tau-\eps))}{\alpha-1}\left(1-\frac{1}{\alpha}\right)^\alpha   
\end{equation}
is achievable in the approximate $(\epsilon, \delta)$-DP framework. By taking the infimum over all $\alpha \in (1, \infty)$, a tighter achievable bound on $\delta(\eps)$ is obtained.\footnote{Note that this only gives an achievable bound on $\delta(\eps)$. This is due to a bounding step in the proof of Proposition 12 in \cite{NEURIPS2020_b53b3a3d} to upper bound $\mathbb{E}[\max\{0, 1-\exp(\eps-L_{M,x,x'}(Y))\}] \leq c\exp((\alpha-1)\tau)$, where $L_{M,x,x'}(Y)$ is the privacy loss random variable defined in Section \ref{sec:prelim}.}

For $\tau$ in \eqref{eq:delta:bound}, we can use the analytical results for the R\'enyi divergence of the OSGT mechanism derived in Appendix \ref{app_renyi}. Specifically, we use \eqref{eq:renyi4}, which is repeated here:
$$
\tau = \frac{\alpha\Delta_2^2}{2\sigma^2} +\frac{k}{\alpha-1} \log\left(\frac{\sqrt{2\pi\sigma^2}}{S}\overline{B}\right),$$
where $\overline{B}$ was given in \eqref{eq:renyi3}. Using this $\tau$ gives an achievable $\delta(\eps)$ for the OSGT mechanism as
\begin{equation}\label{eq:deltaosgtbound}
 \delta^{\mathcal T}(\eps) = \left(\frac{\sqrt{2\pi\sigma^2}}{S}\overline{B}\right)^k\frac{\exp((\alpha-1)(\frac{\alpha\Delta_2^2}{2\sigma^2}-\eps))}{\alpha-1}\left(1-\frac{1}{\alpha}\right)^\alpha.
\end{equation}

\begin{figure}
         \centering
         \includegraphics[width=8.8cm]{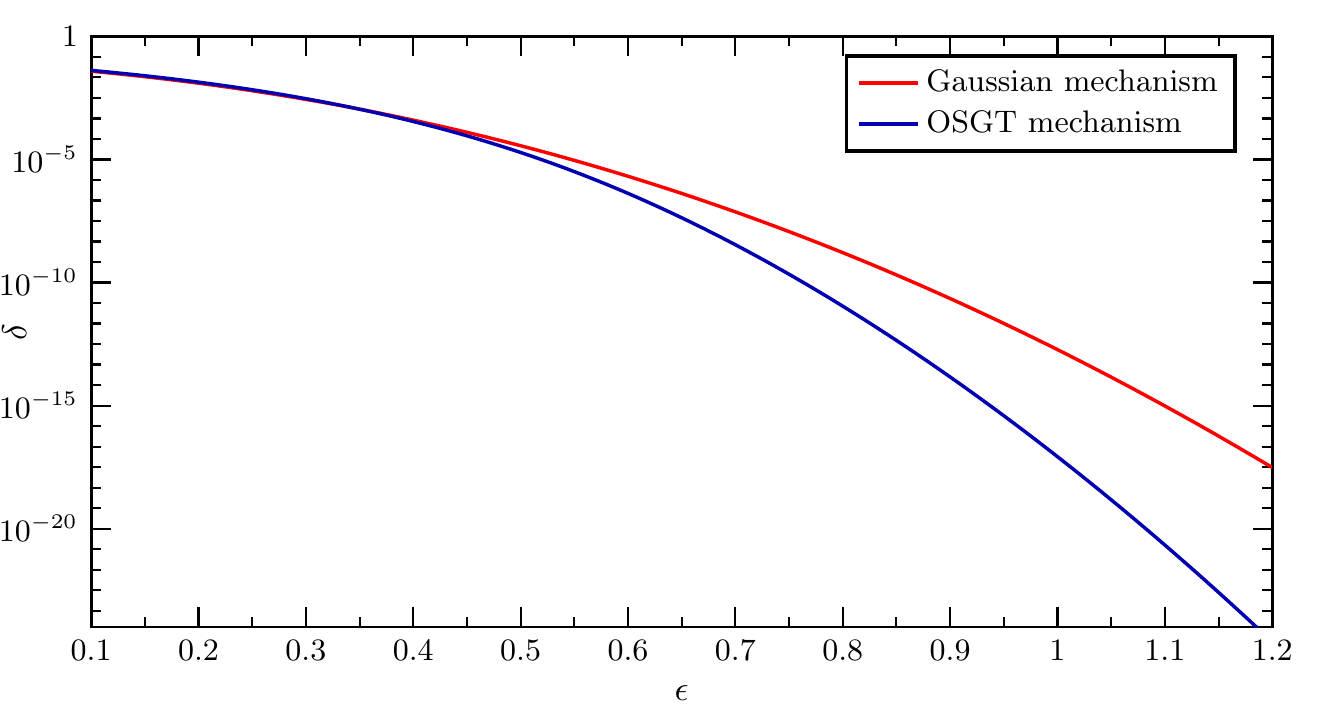}
     \caption{The conversion of concentrated DP of the OSGT and Gaussian mechanisms through minimizing \eqref{eq:deltaosgtbound} and \eqref{eq:deltagaussbound} for $k = 8$, $m=15$, $\sigma^2 = 630$ and $\sigma_g^2 = V(m,\sigma^2) \approx 400$.} \label{fig:zcdp2delta}
 \end{figure}

In contrast, an achievable $\delta$ for the Gaussian mechanism is
\begin{equation}\label{eq:deltagaussbound}
 \delta^{\N}(\eps) = \frac{\exp((\alpha-1)(\alpha\frac{\Delta_2^2}{2\sigma_g^2}-\eps))}{\alpha-1}\left(1-\frac{1}{\alpha}\right)^\alpha.
\end{equation}
To compare the achievable $\delta$ for the OSGT and Gaussian mechanisms, we conduct the following experiment. We set $k = 8$, $m=15$ and $\sigma^2 = 630$. We also set $\Delta_2^2 = k$. This can represent a $k$-dimensional counting query, where the sensitivity of each counting query is $\Delta = 1$. In the case of the OSGT mechanism, an i.i.d. $Y_i \leftarrow \OSGT$ with $V(m,\sigma^2) \approx 400$ is added to each query component. In the case of the Gaussian mechanism, $Y_i \leftarrow \N(0, \sigma^2_g)$ with $\sigma^2_g = V(m,\sigma^2) \approx 400$ is added to each query component. We used the arbitrary precision Python library \texttt{mpmath} \cite{mpmath} to compute $\overline B$ in \eqref{eq:renyi3} (to get reliable results for the last term in \eqref{eq:renyi3} corresponding to the integral $I_2$ specified in Appendix \ref{app_renyi}). The best respective $\alpha$'s minimizing \eqref{eq:deltaosgtbound} and \eqref{eq:deltagaussbound} are numerically found for each target $\eps$. The results are shown in Fig. \ref{fig:zcdp2delta}, where we can observe that even for multi-dimensional queries, the OSGT mechanism can provide a better privacy performance. For example, at $\eps = 0.9$, $\delta^{\mathcal T}(0.9) \approx 1.44\times 10^{-14}$, which is three orders of magnitude smaller than $\delta^{\mathcal N}(0.9) \approx 2.23\times 10^{-11}$.   
 
\section{Conclusions}\label{sec:conclusion}
In this paper, we proposed and studied a new output perturbation differential privacy mechanism called offset symmetric Gaussian tails (OSGT) mechanism, where the noise distribution is obtained through using the normalized tails of two symmetric Gaussian distributions at locations $\pm m$ and with scale $\sigma$. The OSGT distribution decays in sub-Gaussian manner for sufficiently small ratios $m/\sigma$, which is desirable for reducing outliers when responding to queries. Crucially, the variance of the OSGT distributed mechanism is strictly smaller than its input design parameter $\sigma^2$, while its privacy performance is dictated by $\sigma^2$ (as well as by $m$, 2-sensitivity $\Delta_2$, and the query dimension $k$). Therefore, when comparing the OSGT mechanism and Gaussian mechanism at the same effective variance, the OSGT mechanism can provide better differential privacy protection compared to the Gaussian mechanism, as we measured and demonstrated through the approximate $(\epsilon, \delta)$ and zCDP frameworks.

One of the challenges in deriving the privacy performance of the OSGT mechanism is the combination of 1-norm and 2-norm in its distribution. We circumvented this through theoretical bounding techniques for its zCDP and $\delta(\eps)$ performance. As future work, it would be interesting to design experiments on real or synthetic datasets, where one applies the multi-dimensional OSGT mechanism and measures (on the output) the actual resulting $\delta(\eps)$ performance based on the outcomes. We finally remark that due to its ``Gaussian origin", the exact sampling method for the discrete Gaussian presented in \cite{NEURIPS2020_b53b3a3d} can be easily adapted to sample exactly a discrete OSGT distribution (simply through rejecting values between $[-m:m-1]$. We leave a full treatment of the discrete OSGT distribution for future work.

\appendices
\section{Proof of Proposition \ref{prop:tailrel}}\label{app_Gaustail}
\begin{proof}[\unskip\nopunct]
Let the scale of the Gaussian distribution be $\sigma_g = \sqrt{V(m, \sigma^2)}$ according to \eqref{eqvar}. Using Definition \ref{def:tail2}, we have  
\begin{equation*}
    \begin{split}
     &\lim_{y\rightarrow \infty}\frac{f_{1}\left ( y \right )}{f_{2}\left ( y \right )}=\mathrm{exp}\left ( \lim_{y\rightarrow \infty}\left ( \mathrm{log} f_{1}\left ( y \right )-\mathrm{log}f_{2}\left ( y \right )  \right ) \right)\\
    &=\mathrm{exp}\left ( \lim_{y\rightarrow \infty}\left ( \mathrm{log} {\frac{1}{S}}e^{-\frac{\left ( y+m \right )^{2}}{2\sigma^{2}}}-\mathrm{log}\frac{1}{\sqrt{2\pi\sigma_g^2}}e^{-\frac{y^{2}}{2{\sigma}_g^{2}}}\right ) \right)\\
    &=\mathrm{exp}\left ( \lim_{y\rightarrow \infty}\left (-\frac{\left ( y+m \right )^{2}}{2\sigma^{2}}+\frac{y^{2}}{2{\sigma}_g^{2}}+C\right ) \right)\\
    &=\mathrm{exp}\left ( \lim_{y\rightarrow \infty}\left (\frac{2\left ( \sigma^{2}-{\sigma}_g^{2} \right )y^{2}-4{\sigma}_g^{2}my-2{\sigma}_g^{2}m}{4\sigma^{2}{\sigma}_g^{2}}+K\right ) \right)\\
    &= \infty,
    \end{split}
\end{equation*}
where $K$ is a constant and $ \sigma^{2}-{\sigma}_g^{2}>0$ according to Proposition \ref{prop:var}.
\end{proof}

\section{Proof of Proposition \ref{prop:4}}\label{app_Laptail}
\begin{proof}[\unskip\nopunct]
  \begin{equation}
   \begin{split}
    &\lim_{y\rightarrow \infty}\frac{f_{1}\left ( y \right )}{f_{2}\left ( y \right )}=\exp\left ( \lim_{y\rightarrow \infty}\left ( \log f_{1}\left ( y \right )-\log f_{2}\left ( y \right )  \right ) \right)\\
    &=\exp\left ( \lim_{y\rightarrow \infty}\left ( \log {\frac{1}{2\lambda
    }}e^{-\frac{|y|}{\lambda}}-\log {\frac{1}{S}}e^{-\frac{\left ( y+m \right )^{2}}{2\sigma^{2}}}\right ) \right)\\
    &=\exp\left ( \lim_{y\rightarrow \infty}\left (\frac{-y}{\lambda}+\frac{\left ( y+m \right )^{2}}{2\sigma^{2}}+C\right ) \right)\\
    &=\exp\left ( \lim_{y\rightarrow \infty}\left (\frac{\lambda y^{2}+2\left (\lambda m-\sigma^{2}  \right )y+m^{2}} {2\sigma^{2}\lambda}+C\right ) \right)\\
    &=\infty.
        \end{split}
\end{equation}
\end{proof}

\section{Proof of Theorem \ref{thm:delta}}\label{app_delta1D}
\begin{proof}[\unskip\nopunct]
First, using \eqref{eq:osgt1case} and upon cancelling constant terms, $\log$'s and $\exp$'s, we write the privacy loss function as 
\begin{equation*}
    \begin{split}
     \ell_{M,x,x'}(y)&=\log\left(\frac{f_{M(x)}(y)}{f_{M(x')}(y)}\right)\\
     &= \frac{(y-q')^2}{2\sigma^2}\!-\! \frac{(y-q)^2}{2\sigma^2}\!+\! \frac{m|y-q'|}{\sigma^2}\!-\!\frac{m|y-q|}{\sigma^2}.
    \end{split}
\end{equation*}
Therefore, the set $E^*$ reduces to 
\begin{equation}
    \begin{split}
     E^* &= \{y: \ell_{M,x,x'}(y) \geq \eps \}\\
     &= \{y: \log(f_{M(x)}(y)/f_{M(x')}(y)) \geq \eps \}\\
     &=\{y: (y-q')^2 - (y-q)^2\\
     &\hspace{2em}+ 2m|y-q'|-2m|y-q| \geq 2\sigma^2\eps \}.   
    \end{split}
\end{equation}
In the remainder of the proof and without loss of generality, we assume $q<q'$ (otherwise, we will change the labels for $x$ and $x'$ and proceed).

Denote by $\bar{q} \doteq \frac{q+q'}{2}$ the mid point between $q$ and $q'$. We now prove, via establishing contradiction, that in characterizing $E^*$, it suffices to only consider $y \leq \bar{q}$. Assume some $q< \bar{q} <  y < q' $ belongs to $E^*$. Then, we must have $(y-q')^2-(y-q)^2+ 2m|y-q'|-2m|y-q| = -2\Delta_{q,q'}y + q'^2-q^2 + 2m(q'-y-y+q) \geq 2\sigma^2\eps$. This implies $y \leq \bar{q} - \frac{\sigma^2\eps}{2m+\Delta_{q,q'}} < \bar{q}$, which is a contradiction. Similarly, assuming that some $ q' < y$ belongs to $E^*$ leads to a contradiction.
Before proceeding further, we simplify the term $(y-q')^2-(y-q)^2+ 2m|y-q'|-2m|y-q|$ used in the definition of $E^*$ for the two remaining cases:
\begin{enumerate}
    \item For $y \leq q < q'$, we have 
\begin{align}\nonumber
l_1(y) &= (y-q')^2 - (y-q)^2+ 2m|y-q'|-2m|y-q|\\
&= -2\Delta_{q,q'}y + q'^2-q^2 + 2m\Delta_{q,q'}\\
&=-2\Delta_{q,q'}y + 2\Delta_{q,q'}(\bar{q}+ m).
\label{eq:case1}
\end{align}
\item For $q \leq y \leq \bar{q}$, we have 
\begin{align}\nonumber
l_2(y)&= (y-q')^2 - (y-q)^2+ 2m|y-q'|-2m|y-q|\\\nonumber&= -2\Delta_{q,q'}y + q'^2-q^2 -4my+4m\bar{q}\\&=-2y(\Delta_{q,q'}+2m)+2\bar{q}(\Delta_{q,q'}+2m),\label{eq:case2}
\end{align}
\end{enumerate}
where both are decreasing functions of $y$. Now we check whether it is possible for $q$ (the query output on dataset $x$) to belong to $E^*$:
\begin{equation}\label{eq:qbelongE}
    \begin{split}
    q \in E^* &\quad \Leftrightarrow \quad (q-q')^2 + 2m|q-q'| \geq 2\sigma^2\eps\\  
    &\quad \Leftrightarrow \quad \frac{\sigma^2\eps}{\Delta_{q,q'}} \leq \frac{\Delta_{q,q'}}{2}+m.    
    \end{split}
\end{equation}

In other words, for ``small enough" $\eps$, we will have $q \in E^*$. We now prove that if $q \in E^*$, then all $y \leq q$ must also belong to $E^*$. Recall that $l_1(y)$ in \eqref{eq:case1} is a decreasing function of $y \leq q$. Therefore, if $y = q \in E^*$, it means that $l_1(q) \geq 2\sigma^2\eps$. Therefore, all $y \leq q$ also satisfy $l_1(y) \geq 2\sigma^2\eps$ and hence, belong to $E^*$. Therefore, when $q \in E^*$, we need to find the value of $y^* \in [q, \bar{q}]$, for which $E^* = \{y: y \leq y^*\}$. Referring to \eqref{eq:case2}, 
\begin{equation}
    \begin{split}
     l_2(y) &= -2y(\Delta_{q,q'}+2m)+2\bar{q}(\Delta_{q,q'}+2m) \geq 2\eps\sigma^2 \\
&\quad \Leftrightarrow \quad 
y\leq y^* = \bar{q}-\frac{\eps\sigma^2}{\Delta_{q,q'}+2m}.   
    \end{split}
\end{equation}

To summarize, 
\begin{align}\label{eq:ystar1}
\frac{\sigma^2\eps}{\Delta_{q,q'}} \leq \frac{\Delta_{q,q'}}{2}+m \quad \Leftrightarrow \quad E^* = \left\{y: y \leq \bar{q}-\frac{\eps\sigma^2}{\Delta_{q,q'}+2m}\right\}.
\end{align}

Now, we assume ${\sigma^2\eps}/{\Delta_{q,q'}} >{\Delta_{q,q'}}/{2}+m$ and aim to find the set $E^*$. In this case and according to \eqref{eq:qbelongE}, $q\notin E^*$. Since, both $l_1$ and $l_2$ are decreasing functions of $y$, no $y$ greater than $q$ can belong to $E^*$. Therefore, we will find the value of $y^{*} \in (-\infty, q)$, for which $E^* = \{y: y \leq y^{*}\}$.
\begin{equation}
    \begin{split}
     l_1(y) &= -2\Delta_{q,q'}y + 2\Delta_{q,q'}(\bar{q}+ m) \geq 2\eps\sigma^2\\
&\quad \Leftrightarrow \quad y\leq  y^{*} = \bar{q}+m-\frac{\eps\sigma^2}{\Delta_{q,q'}} < q,   
    \end{split}
\end{equation}
where the last equality follows from the assumption ${\sigma^2\eps}/{\Delta_{q,q'}} >{\Delta_{q,q'}}/{2}+m$ and is consistent with the derivations. To summarize, 
\begin{align}\label{eq:ystar2}
\frac{\sigma^2\eps}{\Delta_{q,q'}} >\frac{\Delta_{q,q'}}{2}+m \quad \Leftrightarrow \quad E^* = \left\{y: y \leq \bar{q}+m-\frac{\eps\sigma^2}{\Delta_{q,q'}}\right\}.
\end{align}
Following the results in \cite{balle18a}, we can determine $\delta$ for the two OSGT mechanisms $M(x)$ and $M(x')$ as the solution to the following integral: 
\begin{align}
\delta &= \int_{E^*} (f_{M(x)}(y)-e^\eps f_{M(x')}(y))dy\\=&
\int_{-\infty}^{y^*}\frac{1}{S'}\exp\left({-\frac{(y-q)^2}{2\sigma^{2}}}\right)\exp\left(-\frac{m|y-q|}{\sigma^{2}}\right)dy\\
&\quad - e^\eps \int_{-\infty}^{y^*} \frac{1}{S'}\exp\left({-\frac{(y-q')^2}{2\sigma^{2}}}\right)\exp\left(-\frac{m|y-q'|}{\sigma^{2}}\right)dy\\
&=\int_{-\infty}^{y^*-q} \frac{1}{S'}\exp\left({-\frac{u^2}{2\sigma^{2}}}-\frac{m|u|}{\sigma^{2}}\right)du\\
&\quad - e^\eps \int_{-\infty}^{y^*-q'} \frac{1}{S'}\exp\left({-\frac{u'^2}{2\sigma^{2}}}-\frac{m|u'|}{\sigma^{2}}\right)du',
\end{align}
where we have used change of variables $u = y-q$ in the first integral and $u' = y-q'$ in the second integral. The value of $y^*$ was derived in \eqref{eq:ystar1} and \eqref{eq:ystar2}, respectively for the two cases where ${\sigma^2\eps}/{\Delta_{q,q'}}$ is smaller or greater than ${\Delta_{q,q'}}/{2}+m$. Now we can use the cdf of a zero-mean OSGT random variable given in \eqref{cdf} to simplify $\delta$ as 
\begin{align}
\delta &= F^{\mathcal{T}}(y^*-q)-e^\eps F^{\mathcal{T}}(y^*-q').
\end{align}

When ${\sigma^2\eps}/{\Delta_{q,q'}} \leq {\Delta_{q,q'}}/{2}+m$, we get from \eqref{eq:ystar1} that $y^*-q = \frac{\Delta_{q,q'}}{2}-\frac{\eps\sigma^2}{\Delta_{q,q'}+2m} \geq 0$. Therefore, assuming  $m$, $\sigma^2$, and $\eps$ are fixed, we will have
\begin{align}\nonumber
     \delta_1(\Delta_{q,q'}) &=      1\!-\!\frac{1}{2Q\left({m}/{\sigma}\right)}\biggl(Q\left(\frac{2m\!+\!\Delta_{q,q'}}{2\sigma}\!-\!\frac{\eps\sigma}{\Delta_{q,q'}+2m}\right) \\ \label{eq:delta1}
     &\quad\!+\! e^\eps Q\left(\frac{2m\!+\!\Delta_{q,q'}}{2\sigma}\!+\!\frac{\eps\sigma}{\Delta_{q,q'}\!+\!2m}\right)\biggr).  
\end{align}

When $\frac{\sigma^2\eps}{\Delta_{q,q'}} > \frac{\Delta_{q,q'}}{2}+m$, we have from \eqref{eq:ystar2} that $y^*-q = \frac{\Delta_{q,q'}}{2}+m-\frac{\eps\sigma^2}{\Delta_{q,q'}} < 0$. Therefore, assuming  $m$, $\sigma^2$, and $\eps$ are fixed, we will have
\begin{equation}\label{eq:delta2}
    \begin{split}
    \delta_2(\Delta_{q,q'}) &=       \frac{1}{2Q\left({m}/{\sigma}\right)}\biggl(Q\left(\frac{\eps\sigma}{\Delta_{q,q'}}-\frac{\Delta_{q,q'}}{2\sigma}\right)\\
    &\hspace{7em}- e^\eps Q\left(\frac{\Delta_{q,q'}}{2\sigma}+\frac{\eps\sigma}{\Delta_{q,q'}}\right)\biggr).    
    \end{split}
\end{equation}

As the final step, we need to show that both functions $\delta_1$ and $\delta_2$ in \eqref{eq:delta1} and \eqref{eq:delta2} are increasing functions of $\Delta_{q,q'}$ and also continuous at the boundary point  $\frac{\sigma^2\eps}{\Delta_{q,q'}} = \frac{\Delta_{q,q'}}{2}+m$. Therefore, they assume their worst-case value at $\Delta = \sup_{x,x'\in \X^n: x\sim x'} |q'-q|$.
For the function $\delta_2$, we use the results in \cite{balle18a}. Upon inspection of $\delta_2$ in \eqref{eq:delta2}, we conclude that
\begin{equation}
    \begin{split}
     \delta_2(\Delta_{q,q'})&=       \frac{1}{2Q\left({m}/{\sigma}\right)}\biggl(\Prob[\N(\mu, 2\mu) \geq \eps] \\
     &\hspace{7em}- e^\eps \Prob[\N(\mu, 2\mu) \leq -\eps]\biggr),   
    \end{split}
\end{equation}
for $\mu = \frac{\Delta_{q,q'}^2}{2\sigma^2}$, which was shown in \cite{balle18a} to be an increasing function of $\mu$ (and hence $\Delta_{q,q'}$).
For the function $\delta_1$, we write 
\begin{equation}
    \begin{split}
     \delta_1(\Delta_{q,q'}) &=       1-\frac{1}{2Q\left({m}/{\sigma}\right)}\biggl(\Prob[\N(\rho, 2\rho) \leq \eps]\\
     &\hspace{8em}+ e^\eps \Prob[\N(\rho, 2\rho) \leq -\eps]\biggr)   
    \end{split}
\end{equation}
for $\rho = \frac{(2m+\Delta_{q,q'})^2}{2\sigma^2}$. Using similar techniques as in \cite{balle18a}, we can prove that $$g(\rho) = \Prob[\N(\rho, 2\rho) \leq \eps] + e^\eps \Prob[\N(\rho, 2\rho) \leq -\eps]$$ is a decreasing functions of $\rho$ (and hence $\Delta_{q,q'}$). Therefore, $\delta_1(\rho) = 1-g(\rho)/{(2Q\left({m}/{\sigma}\right))}$ is an increasing function of $\rho$ (and hence $\Delta_{q,q'})$. To show this, we write:
\begin{align}
g(\rho) & = \Prob[\N(\rho, 2\rho) \leq \eps] + e^\eps \Prob[\N(\rho, 2\rho) \leq -\eps]\\\nonumber
& = \frac{1}{\sqrt{2\pi}}\int_{-\infty}^{a(\rho)} e^{-t^2/2}dt + e^\eps \frac{1}{\sqrt{2\pi}}\int_{-\infty}^{b(\rho)} e^{-t^2/2}dt,
\end{align}
where $a(\rho) = \eps/\sqrt{2\rho}-\sqrt{\rho/2}$, $b(\rho) = -\eps/\sqrt{2\rho}-\sqrt{\rho/2}$, and  $-b(\rho)^2/2 +\eps= -a(\rho)^2/2$. It can be shown that 
\begin{align*}
\frac{d}{d\rho}g(\rho) & = e^{-a(\rho)^2/2}\left(\frac{-\eps}{\sqrt{8\rho^3}}-\frac{1}{\sqrt{8\rho}}+\frac{-\eps}{\sqrt{8\rho^3}}-\frac{1}{\sqrt{8\rho}}\right) < 0.
\end{align*}
It only remains to show that $\delta_1$ and $\delta_2$ are continuous at the boundary point $\frac{\sigma^2\eps}{\Delta_{q,q'}} = \frac{\Delta_{q,q'}}{2}+m$. From \eqref{eq:delta1}, at  $\eps = \frac{\Delta_{q,q'}^2+2\Delta_{q,q'}m}{2\sigma^2}$, we will have 
\begin{align}\nonumber
\delta_1(\Delta_{q,q'}) &=       1\!-\!\frac{1}{2Q\left({m}/{\sigma}\right)}\left(Q\left({m}/{\sigma}\right)\!+\! e^\eps Q\left({(m\!+\!\Delta_{q,q'})}/{\sigma}\right)\right)\\\label{eq:delta1boundary}
&=0.5-e^\eps\frac{ Q\left({(m+\Delta_{q,q'})}/{\sigma}\right)}{2Q\left({m}/{\sigma}\right)}.
\end{align}
From \eqref{eq:delta2}, at  $\eps = \frac{\Delta_{q,q'}^2+2\Delta_{q,q'}m}{2\sigma^2}$, we will have 
\begin{align}\nonumber
\delta_2(\Delta_{q,q'}) &=       \frac{1}{2Q\left({m}/{\sigma}\right)}\left(Q\left({m}/{\sigma}\right) \!-\! e^\eps Q\left({(m\!+\!\Delta_{q,q'})}/{\sigma}\right)\right)\\\label{eq:delta2boundary}
&=0.5-e^\eps\frac{ Q\left({(m+\Delta_{q,q'})}/{\sigma}\right)}{2Q\left({m}/{\sigma}\right)},
\end{align}
which confirms the continuity of these two functions.
\end{proof}

\section{Proof of Theorem \ref{thm:zcdp}}\label{app_zcdp}
\begin{proof}[\unskip\nopunct]
The R\'enyi divergence of $k$-dimensional OSGT mechanism is:
\begin{equation}\label{eq:renyi}
    \begin{split}
     &D_{\al}(M(x)\lVert M(x'))  = \\
     &\frac{1}{\al-1}\log\int_{y \in \Y} (f_{M(x)}^\mathcal T(y))^\al (f_{M(x')}^\mathcal T(y))^{(1-\al)} dy,   
    \end{split}
\end{equation}
where $\Y = \R^k$ and the integral is a $k$-dimensional one. Referring to Definition \ref{def:osgt:mech}, let us focus on the integral term in the above and rewrite it as
\begin{equation}\nonumber
    \begin{split}
     \frac{1}{S'^k}&\int_{\R^k}  
\exp\left({\frac{-\alpha\norm{y-q}_2^2-(1-\alpha)\norm{y-q'}_2^2}{2\sigma^{2}}}\right) \\
&\quad{}. \exp\left({\frac{-\alpha m\norm{y-q}_1-(1-\alpha) m\norm{y-q'}_1}{\sigma^{2}}}\right)dy.   
    \end{split}
\end{equation}
Let us now study and simplify each  exponential term individually. Noting that $\alpha > 1$,  $\alpha - 1 > 0$, $|a|-|b| \leq |a-b|$ and $\norm{a}_1-\norm{b}_1 \leq \norm{a-b}_1$, we simplify and upper bound the difference of the $1$-norms in the above as
\begin{equation}\label{eq:1norm:bound}
    \begin{split}
    &{\frac{-\alpha m\norm{y-q}_1-(1-\alpha) m\norm{y-q'}_1}{\sigma^{2}}}\\
    &={\frac{- m\norm{\alpha y-\alpha q}_1+ m\norm{(\alpha-1)y-(\alpha-1)q'}_1}{\sigma^{2}}} \\
&\leq {\frac{m\norm{y-A}_1}{\sigma^{2}}},    
    \end{split}
\end{equation}
where $A \doteq (1-\alpha)q'+\alpha q$ is constant for given $\alpha, q$, and $q'$. 

Now note that queries are real-valued and we can write
\begin{equation}\nonumber
    \begin{split}
    -\alpha\norm{y-q}_2^2 &= -\alpha\norm{y}_2^2+2\langle y, \alpha q \rangle  -\alpha\norm{q}_2^2,\\
     -(1-\alpha)\norm{y-q'}_2^2 &= -(1-\alpha)\norm{y}_2^2\\
     &\quad+2\langle y, (1-\alpha) q' \rangle  -(1-\alpha)\norm{q'}_2^2.   
    \end{split}
\end{equation}
Therefore, 
\begin{equation}\nonumber
    \begin{split}
     &\frac{-\alpha\norm{y-q}_2^2-(1-\alpha)\norm{y-q'}_2^2}{2\sigma^{2}}\\ 
     &=\frac{-\norm{y}_2^2+2\langle y, A \rangle -\alpha \norm{q}_2^2-(1-\alpha)\norm{q'}_2^2}{2\sigma^{2}}\\&=
\frac{-\norm{y-A}_2^2 + \norm{A}_2^2-\alpha \norm{q}_2^2-(1-\alpha)\norm{q'}_2^2}{2\sigma^{2}} \\&=\frac{-\norm{y-A}_2^2 + \alpha(\alpha-1){\norm{q-q'}_2^2}}{2\sigma^{2}},   
    \end{split}
\end{equation}
where $A$ is the same as above and $\Delta_{q,q'} \doteq \norm{q-q'}_2$. Given the above, the zCDP is upper bounded by
\begin{equation}\nonumber
    \begin{split}
       & D_{\al}(M(x)\lVert M(x')) \\ 
       &\leq \frac{1}{\al-1}\log\frac{1}{S'^k}\int_{\R^k}  \exp\biggl(\frac{-\norm{y-A}_2^2 + \alpha(\alpha-1){\Delta_{q,q'}}^2}{2\sigma^{2}}\\
       &\hspace{11em}+\frac{m\norm{y-A}_1}{\sigma^{2}}\biggr)dy\\
&=\alpha\frac{\Delta_{q,q'}^2}{2\sigma^2}\\ &\quad{}+\frac{1}{\al-1}\log\frac{1}{S'^k}\int_{\R^k}  \!\!\exp\left(\frac{-\norm{y\!-\!A}_2^2}{2\sigma^{2}}\!+\!\frac{m\norm{y\!-\!A}_1}{\sigma^{2}}\right)dy.
    \end{split}
\end{equation}
Therefore, it remains to simplify the integral above. To do this, we apply the change of variables $u_i = y_i - A_i = y_i - (1-\alpha)q_i'-\alpha q_i$ for $i \in [k]$ to get
        \begin{align}
        &\int_{\R^k}  \exp\left(\frac{-\norm{y-A}_2^2}{2\sigma^{2}}+\frac{m\norm{y-A}_1}{\sigma^{2}}\right)dy \\
        &= 
\int_{\R^k}  \exp\left(\frac{-\norm{u}_2^2}{2\sigma^{2}}+\frac{m\norm{u}_1}{\sigma^{2}}\right)du\\&=
2^k \int_{\R_+^k}  \exp\left(\frac{-\sum_{i=1}^k u_i^2}{2\sigma^{2}}+\frac{m\sum_{i=1}^k u_i}{\sigma^{2}}\right)du\label{eq:symmetry}\\&=
2^k \int_{\R_+^k}  \exp\left(\frac{-\sum_{i=1}^k (u_i-m)^2}{2\sigma^{2}}\right)\exp\left(\frac{km^2}{2\sigma^2}\right)du\label{eq:completesq}\\&=
2^k\exp\left(\frac{km^2}{2\sigma^2}\right)(\sqrt{2\pi\sigma^2})^k(1-Q(m/\sigma))^k.\label{eq:qk}
    \end{align}
where \eqref{eq:symmetry} follows from the symmetry of the integral over $\R^k$. Hence, it suffices to consider the integral over the non-negative section $\R_+^k$. Equation \eqref{eq:completesq} follows from completing the squares and \eqref{eq:qk} follows from the separability of integrals and the definition of the Gaussian $Q$-function (for the negative value $-m/\sigma$). Recalling the definition $S' = 2\sqrt{2\pi\sigma^2} \exp({{m^{2}}/{(2\sigma ^{2})}})Q({m}/{\sigma})$ yields the final result:
\begin{align*}
D_{\al}(M(x)\lVert M(x'))  &\leq \alpha\frac{\Delta_{q,q'}^2}{2\sigma^2} +\frac{k}{\al-1}\log\left(\frac{1-Q(m/\sigma)}{Q({m}/{\sigma})}\right)\\&\leq
\alpha\frac{\Delta_2^2}{2\sigma^2} +\frac{k}{\al-1}\log\left(\frac{1-Q(m/\sigma)}{Q({m}/{\sigma})}\right),
\end{align*}
where the last inequality is through recalling $\Delta_{q,q'} \leq \Delta_2 = \sup_{x,x'\in \X^n:x\sim x'}\norm{q-q'}_2$.
\end{proof}

\section{Analytical Computation of the R\'enyi Differential Privacy for the OSGT Mechanism}\label{app_renyi}
We first present analytical evaluation of \eqref{eq:renyi} for $k = 1$. We will then discuss how to generalize the result to $k > 1$. Consider two real-valued scalar queries $q=q(x)$ and $q'=q(x')$, which give $M(x) = q(x)+Y$ and $M(x') = q(x')+Y$, where $Y \leftarrow \OSGT$. Without loss of generality assume $q \leq q'$ (otherwise swap their labels and proceed). We write \eqref{eq:renyi} as
\begin{equation}\label{eq:renyi2}
     D_{\al}(M(x)\lVert M(x'))  =
     \frac{1}{\alpha-1}\log\frac{1}{S}\left(I_1+I_2+I_3\right),   
\end{equation}
where using appropriate cases for the two OSGT pdf's that are respectively, location-shifted to $q$ and $q'$, we have
\begin{align*}
    I_1 \!&\doteq\! \int_{-\infty}^{q} \!\exp\left({\frac{-\alpha(y-q-m)^2-(1-\alpha)(y-q'-m)^2}{2\sigma^2}}\right) dy,\\
    I_2 \!&\doteq\! \int_{q}^{q'} \!\exp\left({\frac{-\alpha(y-q+m)^2-(1-\alpha)(y-q'-m)^2}{2\sigma^2}}\right) dy,\\
    I_3 \!&\doteq\! \int_{q'}^{\infty} \!\exp\left({\frac{-\alpha(y-q+m)^2-(1-\alpha)(y-q'+m)^2}{2\sigma^2}}\right) dy.\end{align*}
In $I_1$, we apply the change of variable $u = y-q-m$ to get 
\begin{align}
    I_1 = \int_{-\infty}^{-m} \exp\left({\frac{-\alpha u^2-(1-\alpha)(u-\Delta_{q,q'})^2}{2\sigma^2}}\right) du,
\end{align}
where $0 < \Delta_{q,q'} = q'-q < \Delta$. Upon simplifying and completing the square we get 
\begin{align*}
    I_1 &= e^{\alpha(\alpha-1)\Delta_{q,q'}^2/2\sigma^2}\int_{-\infty}^{-m} e^{\frac{-(u-(1-\alpha)\Delta_{q,q'})^2}{2\sigma^2}} du\\&=\sqrt{2\pi \sigma^2}e^{\alpha(\alpha-1)\Delta_{q,q'}^2/2\sigma^2}\Phi(b_1/\sigma),
\end{align*}
where $\Phi$ is the cdf of $\N(0,1)$ and $b_1 = -m+(\alpha-1)\Delta_{q,q'}$. In $I_3$, we apply the change of variable $u = y-q+m$ to get 
\begin{align*}
    I_3 = \int_{\Delta_{q,q'}+m}^{\infty} \exp\left({\frac{-\alpha u^2-(1-\alpha)(u-\Delta_{q,q'})^2}{2\sigma^2}}\right) du,
\end{align*}
Upon simplifying and completing the square we get 
\begin{align*}
    I_3 &= e^{\alpha(\alpha-1)\Delta_{q,q'}^2/2\sigma^2}\int_{\Delta_{q,q'}+m}^{\infty} e^{\frac{-(u-(1-\alpha)\Delta_{q,q'})^2}{2\sigma^2}} du\\&=\sqrt{2\pi \sigma^2}e^{\alpha(\alpha-1)\Delta_{q,q'}^2/2\sigma^2}\Phi(b_2/\sigma),
\end{align*}
where $b_2 = -m-\alpha\Delta_{q,q'}$. In $I_2$, we apply the change of variable $u = y-q+m$ to get 
\begin{align*}
    I_2 \!=\! \int_{m}^{\Delta_{q,q'}+m} \exp\left(\frac{-\alpha u^2-(1-\alpha)(u-\Delta_{q,q'}-2m)^2}{2\sigma^2}\right) du.
\end{align*}
Upon simplifying and completing the square we get
\begin{align}\nonumber
    I_2 &= e^{\frac{\alpha(\alpha-1)(\Delta_{q,q'}+2m)^2}{2\sigma^2}}\int_{m}^{\Delta_{q,q'}+m}  e^{-\frac{(u-(1-\alpha)(\Delta_{q,q'}+2m))^2}{2\sigma^2}} du
\\\label{eq:I2analytical}&=A\sqrt{2\pi \sigma^2}e^{\alpha(\alpha-1)\Delta_{q,q'}^2/2\sigma^2}\left(\Phi({b_3}/{\sigma})-\Phi({b_4}/{\sigma})\right),
\end{align}
where $$A = e^{\alpha(\alpha-1)(4m\Delta_{q,q'}+4m^2)/2\sigma^2},$$
and 
$$b_3 = \alpha\Delta_{q,q'}-m(1-2\alpha), \quad b_4 = b_3 -\Delta_{q,q'}.$$ Putting all results together, we get 
\begin{align*}
     &D_{\al}(M(x)\lVert M(x')) = \frac{\alpha\Delta_{q,q'}^2}{2\sigma^2} +\frac{1}{\alpha-1} \log\left(\frac{\sqrt{2\pi\sigma^2}}{S}B\right)\\
     &B = \Phi(b_1/\sigma)+\Phi(b_2/\sigma)+A\left(\Phi({b_3}/{\sigma})-\Phi({b_4}/{\sigma})\right).
\end{align*}
Through a linear exhaustive search in $\Delta_{q,q'} \in [0, \Delta]$, we numerically check to ensure $D_{\al}(M(x)\lVert M(x'))$ assumes its worst case at the global query sensitivity $\Delta = \max \Delta_{q,q'}$, for which case we have
\begin{align}\nonumber
     &D_{\al}(M(x)\lVert M(x')) \leq \frac{\alpha\Delta^2}{2\sigma^2} +\frac{1}{\alpha-1} \log\left(\frac{\sqrt{2\pi\sigma^2}}{S}\overline{B}\right),\\\label{eq:renyi3}
     &\overline{B} = \Phi((-m+(\alpha-1)\Delta)/\sigma)\\\nonumber&\qquad +\Phi((-m-\alpha\Delta)/\sigma)+e^{\alpha(\alpha-1)(4m\Delta+4m^2)/2\sigma^2}\times\\\nonumber
     &\qquad\left(\Phi\left(\frac{\alpha\Delta\!-\!m(1\!-\!2\alpha)}{\sigma}\right)\!-\!\Phi\left(\frac{(\alpha\!-\!1)\Delta\!-\!m(1\!-\!2\alpha)}{\sigma}\right)\right).
\end{align}
Now, we specify the result for the $k$-dimensional OSGT mechanism. Note that the mechanism is i.i.d. across different query dimensions. Therefore, the integral in \eqref{eq:renyi} becomes separable, where a similar procedures as above can be followed for each dimension. Let $\Delta_i = \max |q_i-q'_i|$ for dimension $i \in [k]$. We will then have
\begin{align*}
     &D_{\al}(M(x)\lVert M(x')) \!\leq\! \frac{\alpha\Delta_2^2}{2\sigma^2} \!+\!\frac{1}{\alpha\!-\!1} \log\left(\frac{\sqrt{2\pi\sigma^2}^k}{S^k}\prod_{i=1}^k\overline{B_i}\right),
\end{align*}
where each $B_i$ is defined as per \eqref{eq:renyi3} for $\Delta_i$. In particular, if query sensitivities are identical: $\Delta_i = \Delta$, we will have 
\begin{align}\label{eq:renyi4}
     &D_{\al}(M(x)\lVert M(x')) \!\leq\! \frac{\alpha\Delta_2^2}{2\sigma^2} +\frac{k}{\alpha-1} \log\left(\frac{\sqrt{2\pi\sigma^2}}{S}\overline{B}\right).
\end{align}
This completes our derivation.

\bibliographystyle{IEEEtran}
\bibliography{References}
\end{document}